\documentclass[a4,aps,amsmath,floatfix]{revtex4}
\usepackage{graphicx}
\usepackage{color}
\usepackage{enumerate}
\newcommand{\be}{\begin{equation}}
\newcommand{\ee}{\end{equation}}
\newcommand{\ba}{\begin{array}}
\newcommand{\ea}{\end{array}}
\newcommand{\bea}{\begin{eqnarray}}
\newcommand{\eea}{\end{eqnarray}}
\newcommand{\bdm}{\begin{displaymath}}
\newcommand{\edm}{\end{displaymath}}

\begin{document}

\title{Hysteresis in Anti-Ferromagnetic Random-Field Ising Model at Zero 
Temperature}
\author{Lobisor Kurbah}
\affiliation{%
Physics Department \\ North Eastern Hill University \\ 
Shillong-793 022, India}%
\author{Prabodh Shukla}
\email{shukla@nehu.ac.in}
\affiliation{%
Physics Department \\ North Eastern Hill University \\ 
Shillong-793 022, India}%


\begin{abstract}
We study hysteresis in anti-ferromagnetic random-field Ising model at 
zero temperature. The external field is cycled adiabatically between 
-$\infty$ and $\infty$. Two different distributions of the random-field 
are considered, (i) a uniform distribution of width $2\Delta$ centered 
at the origin, and (ii) a Gaussian distribution with average value zero 
and standard deviation $\sigma$. In each case the hysteresis loop is 
determined exactly in one dimension and compared with numerical 
simulations of the model.
\end{abstract}

\maketitle

\section{Introduction}

Hysteresis is a non-equilibrium effect commonly observed in systems 
subjected to a cyclic force ~\cite{bertotti}. It means that the response 
to a changing force depends on the history of the force. In particular, 
the response in increasing force is different from that in decreasing 
force. This is caused by the delay in responding to the force. 
Theoretically hysteresis should disappear if the force changes 
sufficiently slowly but this often corresponds to unrealistically long 
time periods such as the life span of an experimentalist. Several 
complex and disordered systems like permanent magnets show hysteresis 
over the longest practical time scales. Experience with spinglasses and 
other systems containing quenched disorder ~\cite{young} has revealed 
that the free energy landscape of such systems comprises a large number 
of local minima (metastable states). The number of local minima is 
thermodynamically large. The barriers between the local minima are also 
large compared with the thermal energy of the system. Consequently, in 
the absence of a driving field the system gets trapped in one of the 
local minima and is unable to explore the entire phase space over 
practical time scales. In this situation the thermal relaxation time of 
the system $\tau$ is much larger than the relaxation time of its 
constituent units (individual spin-flips) as well as the period 
$2\pi/\omega$ of the cyclic driving field. A useful approximation is to 
assume $\tau$ to be infinite or equivalently the system to be at 
absolute zero temperature. This makes the dynamics of the system 
deterministic and more amenable to analytic solutions and simulations 
without compromising the essential physics of the problem. We take the 
limit $T \rightarrow 0$ before the limit $\omega \rightarrow 0$ to 
obtain nonvanishing hysteresis in the limit $\omega=0$.

In an extensive and pioneering work Sethna et al ~\cite{sethna1,sethna2, 
dahmen} used the random-field Ising model ~\cite{imryma} along with the 
Glauber dynamics ~\cite{glauber} at zero temperature to study hysteresis 
in ferromagnets with quenched disorder. They analyzed their model using 
numerical simulations, mean field theory, Wilson's renormalization group 
~\cite{wilson}, and compared it with experiments. Their model reproduces 
several experimentally observed features. These include familiar shapes 
of hysteresis loops, Barkhausen noise ~\cite{stoner}, and return point 
memory. Interestingly the model predicts the existence of a 
non-equilibrium critical point on each half of the hysteresis loop. This 
is based on a Gaussian distribution of the random-field with mean value 
zero and standard deviation $\sigma$ that plays the role of a tuning 
parameter in the model. The model may be solved exactly in one dimension 
and on Bethe lattices of a general coordination number $z$ 
~\cite{shukla1,shukla2,dhar}. Above a lower critical coordination number 
$z$ ~\cite{dhar,sabhapandit1} there is a critical value $\sigma_c$ such 
that for $\sigma < \sigma_c$, each half of the hysteresis loop has a 
first order jump in the magnetization at some applied field $h$. The 
size of the jump goes to zero as $\sigma \rightarrow \sigma_c$ from 
below. If $h_c$ is the critical field at which the jump vanishes, 
$\{h_c,\sigma_c\}$ is a non-equilibrium critical point showing scaling 
of thermodynamic functions and universality of critical exponents in its 
vicinity. This is reminiscent of equilibrium critical phenomena and 
appears to have a fair amount of experimental support in the field of 
hysteresis as well. A generalization of the model 
~\cite{silveira1,silveira2,shukla3,shukla4} to $n$-component ($n>1$) 
classical spins shows the existence of critical points in the 
generalized model as well. The critical exponents of the generalized 
model are in the universality class of the random-field Ising model 
($n=1$) if the critical point occurs at a non-zero value of 
magnetization or the applied field. This is understandable because a 
non-zero value of magnetization or the applied field picks a unique 
direction in the system that effectively reduces its symmetry to that of 
an Ising model. This lends further support to the broad agreement 
between experiments and predictions of RFIM for hysteresis.

Hysteresis in the anti-ferromagnetic random field Ising model 
~\cite{shukla5,shukla6, shukla7,shukla8} has received relatively little 
attention as compared to its ferromagnetic counter part ~\cite{sethna1, 
sethna2,dahmen,dhar,shukla9,shukla10,sabhapandit1,illa1,illa2, 
illa3,sabhapandit2}. This is partly due to the difficulty of obtaining 
analytic solutions in the anti-ferromagnetic case. For the ferromagnetic 
case, exact expressions for the major and minor hysteresis loops have 
been obtained in one dimension as well as on a Bethe lattice of 
coordination number $z$ for a bounded as well as a Gaussian distribution 
of the quenched field ~\cite{dhar,shukla7,shukla8}. The distribution of 
Barkhausen jumps (avalanches) has also been obtained 
~\cite{sabhapandit1}. Several other aspects of the ferromagnetic model 
have been studied in the mean field theory as well as on periodic 
lattices ~\cite{ sabhapandit2,illa1,illa2,illa3}. In the 
anti-ferromagnetic case an expression for the major loop has been 
obtained ~\cite{shukla4} in one dimension in case the quenched field has 
a uniform distribution of width $\Delta$ centered at the origin and 
$\Delta < |J|$ where $J$ is the anti-ferromagnetic exchange interaction. 
The purpose of the present paper is to extend this result to $\Delta \ge
|J|$ as well. The results presented here are also applicable to 
unbounded distributions of the quenched random-field such as the 
Gaussian distribution.

It may appear rather surprising at first sight that there should be any 
difficulty in solving a one-dimensional Ising model at zero temperature. 
The difficulty arises primarily from the presence of quenched random 
fields. Problems with quenched disorder are difficult to analyze 
analytically. Besides this the spin-flip dynamics with 
anti-ferromagnetic interactions is more complicated than its 
ferromagnetic partner. Consider two spin systems of equal size and 
having the same realization of quenched field distribution. Let one 
system have ferromagnetic nearest neighbor interaction $J$ and the other 
an anti-ferromagnetic interaction $-J$. In equilibrium, the ground 
states of the two systems on a bi-partite lattice are related to each 
other by symmetry. Evidently no such relation is available between 
non-equilibrium metastable states of the two systems. An applied field 
$h_a$ increasing adiabatically from $h_a=-\infty$ to $h_a=\infty$ takes 
both systems from a stable state with all spins pointing down (i.e. 
aligned along $h_a=-\infty$) to all spins pointing up. Although the end 
points of the trajectory are the same for both systems but the 
magnetization paths are different. In particular the number of 
metastable states along the two paths are different. In the 
ferromagnetic case, spins tend to flip up in avalanches and do not flip 
down in increasing field. The anti-ferromagnetic dynamics is marked by 
the absence of avalanches. This is because a spin flipping up at an 
applied field $h_a$ prevents its neighbors from flipping up at the same 
field. However, a spin flipping up at $h_a$ occasionally causes its 
neighbor to flip down at $h_a$. This is a kind of a reverse avalanche in 
anti-ferromagnetic dynamics that involves only two spins including the 
spin that triggers the avalanche. The forward avalanches in the 
ferromagnetic case, and the reverse avalanches in the anti-ferromagnetic 
case provide a mechanism for irreversibility in the two models 
respectively and give rise to hysteresis. Due to the smaller size of 
reverse avalanches the area of anti-ferromagnetic hysteresis loop is 
much smaller than the area of ferromagnetic loop. Also the Barkhausen 
noise on the ferromagnetic hysteresis loop that is caused by large 
sporadic avalanches is nearly absent in the anti-ferromagnetic case.

The relative difficulty of analyzing anti-ferromagnetic dynamics comes 
from the fact that it is non-Abelian while the ferromagnetic dynamics is 
Abelian. This means as follows. Consider an unstable system at an 
applied field $h_a$ such that one or more spins are not aligned along 
the net field at their site. We relax the system till it is stable. 
Relaxing the system means checking each spin and flipping it if it is 
not aligned along the net field at its site. It is an iterative process 
because flipping a spin may reverse the sign of the net field at its 
nearest neighbors. We have to continue the relaxation process till each 
spin in the system is stable. A dynamics is called Abelian if the end 
result of the relaxation process does not depend on the order in which 
the spins are relaxed. If the result does depend on the order in which 
the spins are relaxed it is called non-Abelian. Consider two nearest 
neighbor spins which are both down but the net field at their sites is 
positive. If the interaction between the spins is ferromagnetic, the 
spins can be relaxed in any order and the end result would be that both 
spins are turned up. This is because turning a spin up makes the net 
field at its neighbor even more positive so that the neighbor also has 
to be turned up. This is not the case with anti-ferromagnetic 
interactions. Turning a spin up decreases the net field at its neighbor 
and it may decrease it below zero so that the neighbor no longer needs 
to be turned up when relaxed. Thus the end result may be one spin up and 
one down. Which one is up depends on which one was turned up first. The 
anti-ferromagnetic dynamics is therefore non-Abelian. As the stable 
state at the end of the relaxation process depends on the order in which 
the unstable spins are relaxed, we have to choose a protocol for the 
order in which the unstable spins are relaxed. At every step, we choose 
to relax the most unstable spin in the system i.e. the one whose 
flipping would lower the energy of the system the most. Locating the 
most unstable spin at every step of the dynamics is what makes the 
anti-ferromagnetic model more tedious to analyze theoretically as well 
as numerically.

Although the aim of the present study is to find an analytic solution of 
a non-equilibrium problem with quenched disorder, we may mention some 
connection with experiments. Relaxation dynamics of any complex 
statistical system belongs to one of two broad categories: (i) where 
relaxation takes place by avalanches, and (ii) where it proceeds by 
single localized events. The ferromagnetic random-field Ising model 
belongs to the category of avalanches. It explains experimental effects 
such as the Barkhausen noise and the possibility of non-equilibrium 
critical points. The anti-ferromagnetic random-field Ising models 
belongs to the second category characterized by the absence of 
avalanches. Due to the absence of avalanches, we do not expect small 
changes in the applied field to cause large changes spanning across the 
system. In other words, we do not expect the response of the system to 
be critical at any value of the applied field. This rules out the 
existence of non-equilibrium critical points in anti-ferromagnets. Our 
calculation shows that the hysteresis loop of an anti-ferromagnet with 
relatively small quenched disorder (to be defined in the following) has 
a wasp-waisted shape i.e. constricted in the middle. In the limit of 
very small disorder the wasp-waisted shape gradually transforms into two 
hysteresis loops joined by a long and narrow region of almost no 
hysteresis. For much larger disorder the familiar pot-belly shape of 
ferromagnetic loops is recovered. Thus the anti-ferromagnets can exhibit 
a wide variety of shapes of hysteresis loops and this feature of our 
model is in general conformity with experiments ~\cite{ayyub,fukuma, 
chiorescu,fullerton,waldmann}. Anti-ferromagnetic hysteresis loops 
comprising three loops are also observed in experiments 
~\cite{takanashi,chew}. This too is understandable if our one-dimensional 
model is extended to lattices with higher coordination number. The 
anti-ferromagnetic model may also apply to other systems that exhibit 
glassy dynamics~\cite{shukla1, kisker,rieger,ritort,toninelli} 
characterized by a single localized events.

The outline of this paper is as follows. In Section II we define the 
model for a general distribution of the quenched random-field centered 
at the origin. We focus on two specific cases: a uniform bounded 
distribution of width $2\Delta$ and a Gaussian distribution of standard 
deviation $\sigma$. If $|J|$ is the magnitude of nearest neighbor 
anti-ferromagnetic interaction, the case $\Delta \le |J|$ is easier to 
treat analytically for reasons to be made clear in the following. The 
case $\Delta > |J|$ is more difficult but comparable to the case of 
unbounded Gaussian distribution. In Section III, we show numerical 
results for the hysteresis loop in three representative cases: (i) 
$\Delta=0.5|J|$, (ii) $\Delta=1.25|J|$, and $\sigma=.5 |J|$. Case (i) 
lies in the ambit of an exact solution obtained earlier 
~\cite{shukla7,shukla8} for $\Delta \le |J|$. In Section IV, we briefly 
review the earlier result because it is needed to proceed to 
distributions with $\Delta >|J|$ and the Gaussian distribution. The 
formalism and presentation of section IV is slightly different from the 
earlier version ~\cite{shukla7,shukla8} on which it is based in order to 
make a smoother transition to the following sections. Sections V, VI and 
VII treat a uniform distribution of arbitrary $\Delta$ and a Gaussian 
distribution of arbitrary $\sigma$. These sections contain the main 
results of this paper. Exact expressions for the hysteresis loops are 
obtained. These have been superimposed on the simulation results shown 
in figures (1)-(6). The fit between the theory and simulations is quite 
good as may be expected from an exact solution. Indeed the two are 
indistinguishable on the scale of the figures. The fact that simulations 
over a relatively small size of the system agree with the exact result 
is due to the super-exponential decay of correlations in this 
system~\cite{evans}. The agreement between simulation and theory also 
justifies (albeit post facto) the implicit assumption in our analysis 
that the system is self-averaging.

\section{The Model}

We consider non-equilibrium anti-ferromagnetic random-field Ising model 
in one dimension at zero temperature. At each site $i$ ($i=1,2,3, 
\ldots,N$) of a linear lattice, there is an Ising spin $ s_{i}=\pm{1}$ 
which interacts with its nearest neighbors through an anti-ferromagnetic 
interaction $J$ ($J<0$). A quenched random-field $h_{i}$ as well as a 
uniform externally applied field $h_a$ acts on $s_i$. The Hamiltonian of 
the system is,

\be H=-J \sum_{i}{s_{i}s_{i+1}}-\sum_{i}h_{i}s_{i}-h_{a}\sum_{i} 
s_{i}\ee

We consider two distributions $\phi(h_i)$ of the random-field 
$\{h_{i}\}$:
 
\begin{enumerate}[(a)]

\item A uniform bounded distribution of width $2\Delta$ centered at the 
origin,

\begin{eqnarray} \phi(h_{i}) & = \frac{1}{2\Delta} &\mbox{if } [-\Delta 
\le h_{i} \le \Delta] \nonumber \\ & = 0, & \mbox{otherwise.} 
\end{eqnarray}

\item A Gaussian distribution with average zero and standard deviation 
$\sigma$,

\be \phi(h_{i}) = \frac{1}{\sqrt{2\pi \sigma^2}}\left. 
e^{-\frac{h_i^2}{2\sigma^2}}.\right. \ee

\end{enumerate}

It is convenient to rewrite $H$ in terms of the net field $f_i$ acting 
on spin $s_i$,

\be H=-\sum_{i}f_is_{i}; \mbox{ }f_{i}= J(s_{i-1}+s_{i+1})+h_{i}+h_{a} 
\ee

The spins $\{s_i(t)=\pm1\}$ obey discrete-time single-spin-flip Glauber 
dynamics at zero temperature i.e. $s_i(t+1)=\mbox{ sign } f_i(t)$. This 
means a spin flips only if it lowers its energy. It also assumes that if 
a spin-flip is allowed, it occurs at a rate $\Gamma$ which is much 
larger than the rate at which the magnetic field $h_{a}$ is varied. Thus 
all flippable spins relax instantly and $s_{i}(t+1)$ has the same sign as 
the net local field $f_{i}(t)$ at its site. 

\be s_{i}(t+1)=\mbox{ sign }{f_{i}(t)} =\mbox{ sign } 
[J\{s_{i-1}(t)+s_{i+1}(t)\}+h_{i}+h_{a}(t)] \ee

Iterative application of the above dynamics leads to a fixed point state 
of the system such that $s_i(t+1)=s_i(t)$ for each spin $s_i(t)$ in the 
system. The condition of adiabatic variation of the applied field or 
equivalently the instant relaxation of spins mentioned previously is 
implemented by holding the applied field constant until a fixed point is 
reached. The fixed point is a local minimum of the energy (metastable 
state) of the system. We denote the fixed point value of $s_i(t)$ by 
$s_i^*$ and characterize the fixed point state by the magnetization 
$m(h_{a})$ per spin,

\be m(h_{a})=\frac{1}{N}\sum_{i}s_i^* \ee

Our aim is to find the magnetization $m(h_a)$ of each metastable state 
visited by the system as the applied field is cycled adiabatically from 
$h_a=-\infty$ to $h_a=\infty$ and back to $h_a=-\infty$. We start with 
$h_{a}=-\infty$ when each spin is necessarily aligned along the applied 
field i.e. we have a fixed point with $m=-1$. Now we increase $h_{a}$ 
slowly till some spin becomes unstable and needs to be flipped. We flip 
this spin and check its neighborhood if any more spins have to be 
flipped. If any neighbors are flipped, we check their neighbors if they 
need to be flipped as well. This process is continued till no more spin 
needs to be flipped i.e. we reach a new fixed point. The applied field 
is held constant during the passage from the old fixed point to the new. 
This procedure is continued up to $h_{a}=+\infty$ when $m=1$. The 
magnetization $m^{u}(h_{a})$ on the upper half of the hysteresis loop 
when $h_{a}$ is slowly decreased from $h_{a}=+\infty$ to $h_{a}=-\infty$ 
can be obtained from $m(h_{a})$ by a symmetry relation 
$m^{u}(h_{a})=-m(-h_{a})$. Therefore the calculation of the lower half 
of the hysteresis loop suffices to determine the entire hysteresis loop.

\section{simulations}

Computer simulations of the preceding model play a useful role in 
guiding its analysis and checking the analytic results. Normally we used 
$10^{3}$ spins on a linear lattice with periodic boundary conditions and 
used $10^{3}$ independent realizations of the random field distribution 
to generate the data. The data was binned in $10^3$ bins in the 
applicable range of the applied field and averaged over different 
realizations of the field distribution. The simulation results shown in 
figures (1)-(6) and some of the other figures in section VIII were 
obtained in this way. It took approximately four hours on our 3 GHz 
desktop to generate the data for each figure. When the estimated 
probability of an event was very small, say of the order of $10^{-6}$, 
we performed the corresponding simulation on a larger system, say $10^6$ 
spins, and a smaller number of independent runs, say $10^2$. This was to 
optimize the accuracy of the data and the time taken to generate it. In 
some cases it was more appropriate to have a larger number of bins. In 
these cases we worked with the data obtained from a large system of 
$10^6$ spins without binning it. We set $J=-1$, and as mentioned 
previously we performed simulations for three cases: (i) $\Delta=0.5$, 
(ii) $\Delta=1.25$, and (iii) $\sigma=0.5$. These values were chosen 
arbitrarily but represent three broad classes of the analytic results.

\subsection{$\Delta=0.5$}

Figure (1) shows the hysteresis loop for $\Delta=0.5$. We see that 
$m(h_a)=-1$ if $h_a\le-2 |J| -\Delta$, and $m(h_a)=1$ if 
$h_a\ge2|J|+\Delta$. The magnetization $m(h_{a})$ rises from -1 to +1 in 
three steps. We call these steps ramp-I ($h_{a}=-2|J|-\Delta$ to 
$h_{a}=-2|J|+\Delta$); ramp-II ($h_{a}=-\Delta$ to $h_{a}=+\Delta$); and 
ramp-III ($h_{a}=2|J|-\Delta$ to $h_{a}=2|J|+\Delta$). The ramps are 
connected to each other by two plateaus; plateau-I ( 
$h_{a}=-2|J|+\Delta$ to $h_{a}=-\Delta$); and plateau-II 
($h_{a}=+\Delta$ to $h_{a}=2|J|-\Delta$). On the plateaus, the 
magnetization remains constant even though the applied field continues 
to increase. Plateaus occur for $\Delta \le |J|$ (small disorder), and 
simulations suggest that magnetization on the plateaus is independent of 
$\Delta$. Numerically, the magnetization on the plateaus is 
approximately $m^{I}=-.135$ on plateau-I, and $m^{II}=.109$ on 
plateau-II. The qualitative shape of $m(h_{a})$ is easy to understand. 
Due to the anti-ferromagnetic interaction between nearest neighbors, 
spins with both neighbors down are the first to turn up in an increasing 
applied field. Such spins turn up on ramp-I. Next are the spins with one 
neighbor up and one down which turn up on ramp-II. Spins with both 
neighbors up require the largest applied field to turn up, and these 
turn up on ramp-III. For $\Delta \le |J|$, the three ramps are well 
separated from each other. In other words, no spin with $n$ up neighbors 
($n=1,2$) can turn up in increasing $h_a$ until all spins with $n-1$ up 
neighbors have turned up. On each ramp, the sequence in which the spins 
turn up is determined by the distribution of the quenched random field. 
Spins with large positive quenched field turn up before spins with a 
lower quenched field. The quenched field lies in the range $-\Delta$ to 
$+\Delta$. Thus each ramp has a width $2\Delta$ along the axis of the 
applied field. When a spin turns up on ramp-I, its nearest neighbors are 
placed in a category so that they cannot turn up before ramp-II. 
Similarly when a spin turns up on ramp-II, its nearest neighbor which is 
down cannot turn up before ramp-III. This is essentially the reason for 
the absence of avalanches in the anti-ferromagnetic RFIM. Occasionally 
on ramp-II and ramp-III, a spin turning up can cause its nearest 
neighbor which is already up to turn down. We may call this a reverse 
avalanche of size unity. There are no long avalanches as in the 
ferromagnetic model. If there were no reverse flips at all there would 
be no hysteresis in the model. The smallness of the reverse flips is the 
reason behind the smallness of the area of the hysteresis loop. In order 
to highlight the separation of the upper and lower halves of the 
hysteresis loop we have plotted in figure (2) the relative separation of 
the two halves relative to their average value. As the majority of the 
spins turn up one at a time, the calculation of $m(h_{a})$ becomes 
essentially a matter of sorting quenched random fields in decreasing 
order on each ramp. The difficulty arises from the fact that {\em{a 
posteriori}} distribution of random fields on unflipped spins that are 
next to a flipped spin is significantly different from the initial 
uniform distribution. The main problem is to calculate this {\em{a 
posteriori}} probability distribution of random fields on unflipped spin 
sites.

\subsection{$\Delta=1.25$}

Figure (3) shows the hysteresis loop for $\Delta=1.25$. We see that the 
three ramps comprising the hysteresis loop in figure (1) have lost their 
individual identity. If $\Delta > |J|$, a spin with one neighbor up can 
turn up on the lower hysteresis loop before all spins with both 
neighbors have turned up. This makes the analysis of the dynamics more 
complex. We shall take it up in Section V and Section VI. Figure (4) is 
a magnified version of figure (3). It shows the relative separation of 
the two halves from their average value at a given applied field.

\subsection{$\sigma=0.5$}

Figure (5) shows the hysteresis loop for $\sigma=0.5$ for a Gaussian 
distribution of the random field. The Gaussian distribution is an 
unbounded distribution. Therefore we may not expect the hysteresis loop 
to comprise of sharp ramps and plateaus as in figure (1). However, 
notice the qualitative similarity between figure (5) for $\sigma=0.5$ 
and figure (1) for $\Delta=0.5$ {\em{sans}} the sharp edges in figure 
(1). The general analysis presented in sections V, VI, VII applies 
equally to a uniform distribution with any value of $\Delta$ and a 
Gaussian distribution for any value of $\sigma$. Figure (6) is a 
magnified version of figure (5) showing the relative separation of the 
two halves of the hysteresis loop from their average value at a given 
applied field. In view of the smallness of the area of the 
anti-ferromagnetic hysteresis loop, the magnified loops are better 
suited for comparing the fit between simulation and the theory to be 
presented in the following sections.

\section{Bounded distribution of quenched field with $\Delta \le |J|$}

If $\Delta \le |J|$ and the applied field $h_a$ is increased 
adiabatically 
from $-\infty$ to $\infty$, the spins with both neighbors down flip up 
first (ramp-I). Next are those with one neighbor up and one down 
(ramp-II). The last category of spins to flip up are those with both 
neighbors up (ramp-III). The ramps are separated by plateau-I and 
plateau-II. The width of the plateaus along the applied field axis 
decreases with increasing $\Delta$ and goes to zero as $\Delta 
\rightarrow |J|$. Magnetization on ramp-I was determined in reference 
~\cite{shukla5} by exploiting a similarity between this problem and the 
problem of random sequential adsorption(RSA)~\cite{evans}. The rate 
equations of the RSA problem were used to determine $m(h_{a})$ on 
ramp-I, but they could not determine $m(h_{a})$ on ramp-II and ramp-III. 
A different approach was introduced in ~\cite{shukla6,shukla7} which 
determined magnetization on all three ramps for $\Delta \le |J|$. We 
recall this approach briefly because it serves as the starting point for 
analyzing magnetization curves for $\Delta > |J|$ as well as for a 
Gaussian distribution of random-fields. The following subsections 
contain the main results obtained in references ~\cite{shukla6,shukla7} 
with some reworking of notation and formalism.

\subsection{Ramp-I}

The analytical results for the magnetization on ramp-I are conveniently 
expressed in terms of three quantities $p_0(h_a)$, $p_1(h_a)$, and 
$p_2(h_a)$. These are the probabilities that a spin which has quenched 
field $h_i$ and which has respectively zero, one, or two nearest 
neighbors up can flip up at applied field $h_a$. Thus,

\be p_n(h_a)=\int_{-2(1-n)|J|-h_a}^{\infty}\phi(h_i) dh_i \mbox{
  }(n=0,1,2)\ee

As the applied field increases adiabatically from $h_a=-\infty$ to 
$h_a=\infty$, spins with both neighbors down begin to flip up at 
$h_a=-2|J|-\Delta$ and continue to flip up till $h_a=-2|J|+\Delta$ at 
which stage ramp-I is completed and there are no down spins whose 
neighbors are also down. The fraction of up spins on ramp-I at an 
arbitrary applied field $h_a$ is given by,

\be P_\uparrow^{I}(h_a)=\frac{1}{2}[1-e^{-2p_0(h_a)}], \ee

The central object in the calculation of $P_\uparrow^{I}(h_a)$ is the 
probability (per site) of finding a pair of adjacent down spins on 
ramp-I at applied field $h_{a}$. We denote this object by the symbol 
$P_{\downarrow\downarrow}^{I}(h_a)$. It is calculated as follows. 
Imagine coloring all sites with $h_{i}+2|J|+h_a \ge 0$ black, and all 
sites with $h_{i}+2|J|+h_a < 0$ white. Consider two adjacent down spins 
A and B shown in Figure (7). The sites A and B can be both white, both 
black, or mixed. Given that A is down, it is clear that the state of B 
can only be influenced by the evolution of the system to the right of B. 
Similarly, given that B is down, the state of A can only be influenced 
by the evolution of the system to the left of A. We shall refer to this 
as the principle of conditional independence~\cite{mityushin}. It 
requires

\be P_{\downarrow\downarrow}^{I}(h_a)=P(A\downarrow|B\downarrow) 
P(B\downarrow|A\downarrow) \ee

where $P(A\downarrow|B\downarrow)$ is the probability that spin at site 
A is down given that spin at B is down, and $P(B\downarrow|A\downarrow)$ 
is the probability that B is down given that A is down. We take up the 
calculation of $P(B\downarrow|A\downarrow)$. If B is a white site, 
$P(B\downarrow|A\downarrow) = 1$ because white sites have not been 
relaxed from their initial state. If B is a black site and the site to 
the right of B is a white site then $P(B\downarrow|A\downarrow)=0$. In 
general $P(B\downarrow|A\downarrow)$ depends on the length of the string 
of black sites to the right of B. Suppose B is a black site, and there 
are $(n-1)$ additional black sites to the right of B. In this case, the 
probability $P^{n}_{B}$ that B is down satisfies the following recursion 
relation,

\be P^{n}_{B}=\frac{1}{n}P^{n-2}_{B}+(1-\frac{1}{n})P^{n-1}_{B} \ee

The rationale for the above recursion relation is as follows. Let the 
black site farthest from B on the right be labeled as the $n$-$th$ 
site. Any of the $n$ sites could flip first. The probability that the 
$n$-$th$ site flips first is therefore equal to $\frac{1}{n}$. If this 
happens, ($n$-1)-$th$ site is prevented from flipping up on ramp-I. The 
probability that B is down is now reduced to the probability that the 
end point of a chain of ($n$-2) black sites is down i.e. $P_{B}^{n-2}$. 
This accounts for the first term in equation (10). The probability that 
$n$-$th$ site is not the first site to flip up is equal to 
$(1-\frac{1}{n})$. Given this situation, the probability that B is down 
is equal to the probability that the end of a string of ($n$-1) black 
sites is down. This accounts for the second term in equation (10). We 
can rewrite the recursion relation (10) as

\be (P^{n}_{B}-P^{n-1}_{B})=-\frac{1}{n}[P^{n-1}_{B}-P^{n-2}_{B}] \ee

It has the solution,

\be P^{n}_{B}=\sum^{n}_{m=0}\frac{(-1)^{m}}{m!} \ee

Summing over various possible values of $n$ with appropriate weight, we 
get

\begin{eqnarray}
P(B\downarrow|A\downarrow)  & = & \sum^{\infty}_{n=0}\sum^{n}_{m=0}
\frac{(-1)^{m}}{m!}{p_0^{n}}(1-p_0) \nonumber \\
&  &   =\sum^{\infty}_{m=0}
\frac{(-1)^{m}}{m!}(1-p_0)\sum^{\infty}_{n=m}p_0^{n}
=\sum^{\infty}_{m=0}\frac{(-p_0)^{m}}{m!}
=e^{-p_0}
\end{eqnarray}
      
In the above array of equations, $p_0$ stands for $p_0(h_a)$. Thus,

\be P_{\downarrow\downarrow}^{I}=e^{-2p_0(h_a)} \ee

Let $P_{\downarrow}^{I}$ be the probability per site of finding
a down spin and $P_{\downarrow\uparrow}^{I}$ the probability
per site of finding a down spin which is followed by an up spin.
Clearly,

\be P_{\downarrow}^{I}=P_{\downarrow\downarrow}^{I} 
+P_{\downarrow\uparrow}^{I}=1-P_{\uparrow}^{I} \ee

Keeping in mind that on ramp-I an up spin must be preceded (as well as 
followed) by a down spin, we get 
$P_{\downarrow\uparrow}^{I}=P_{\uparrow}^{I}$. Thus,

\bdm P_\uparrow^{I}=1-P_{\downarrow\downarrow}^{I}-P_\uparrow^{I} \edm

or,

\be P_\uparrow^{I}=\frac{1}{2}[1-P_{\downarrow\downarrow}^{I}]
               =\frac{1}{2}[1-e^{-2p_0(h_a)}]\ee

The magnetization on ramp-I is given by
\be
m^{I}(h)=2P_\uparrow^{I}(h)-1=-e^{-2p_0(h_a)}
\ee

The exact value of the magnetization on plateau-I is equal to 
$-\frac{1}{e^{2}}$ which is approximately equal to -.135.


\subsection{Plateau-I}

Plateau-I contains down spins in singlets and doublets punctuated by up 
spins. Each spin in a doublet has one neighbor up and one down. 
Therefore the net field on it is simply the sum of the random field 
$h_{i}$ on its site and the applied field $h_{a}$. It turns up when 
$h_i+h_a \ge 0$. The random field lies in the range 
$-\Delta<{h_{i}}<+\Delta$. Therefore an applied field smaller than 
$-\Delta$ is sufficiently negative to pin down all doublets. This 
accounts for the range $(-2|J|+\Delta)<h_a<{-\Delta}$ where the 
magnetization shows a plateau. In each doublet, the spin with the larger 
quenched field $h_{i}$ flips up on ramp-II when $h_i+h_a \ge 0$. The 
spin with the smaller quenched field then becomes a singlet which does 
not flip up before ramp-III. Thus, in order to find the form of ramp-II, 
we need to find the {\em{a posteriori}} distribution of quenched random 
fields on the doublets.

Consider a doublet on plateau-I as shown in Figure (8). The doublet 
sites are denoted as 1 and 2, and the quenched random fields on these 
sites are $h_{1}$ and $h_{2}$. The {\em{a posteriori}} probability 
distributions of $h_{1}$ and $h_{2}$ will be identical by symmetry. 
These distributions $\tilde{\phi}(h_1)$ and $\tilde{\phi}(h_2)$ are 
determined by the relaxation process on ramp-I. If the doublet survives 
up to plateau-I it must exist on ramp-I. Consider ramp-I at an applied 
field $h_a$. Given that site-2 is down at this point, the probability 
that site-1 is also down is equal to $e^{-p_0(h)}$. Site-1 may be down 
because (i) it is a white site i.e. $h_1+2|J|+h_a \le 0$ and therefore 
it could not turn up even if both its nearest neighbors were down, or 
(ii) it is a black site but blocked from turning up by its neighbor that 
has turned up before it. The probability that it is a white site is 
equal to $1-p_0(h_a)$. Therefore the probability that it is a black site 
but down at $h_a$ is equal to $e^{-p_0(h_a)}-\{1-p_0(h_a)\}$. This 
means,

\be Prob(1\downarrow|2\downarrow;h_1+2|J|+h_a\ge 
0)=\int_{-2|J|-h_a}^\Delta \tilde{\phi}(h_1) dh_1 = 
\left[e^{-p_0(h_a)}-\{1-p_0(h_a)\}\right]. \ee

${\tilde{\phi}(h_1)}$ is obtained by taking the derivative of the above 
expression. We get,

\be \tilde{\phi}(h_1) dh_1 = [1-e^{-p_0(-h_1-2|J|)}] \phi(h_1) dh_1 \ee

Assuming $h_1>h_2$, site-1 would turn up on ramp-II when $h_1+h_a=0$. 
The density of sites on ramp-II at this value of the applied field is 
given by,

\be \tilde{\phi}(-h_a) dh_a = [1-e^{-p_0(h_a-2|J|)}] \phi(-h_a) dh_a= 
[1-e^{-p_1(h_a)}] \phi(-h_a) dh_a\ee

We now address an issue which is crucial for determining ramp-II 
correctly. This concerns two adjacent doublets as shown in Figure (9). 
What is the probability per site of observing this object on ramp-I? A 
doublet has an important property. It separates the lattice into two 
parts (one on each side of the doublet) which have evolved uninfluenced 
by each other. Thus, we can separate Figure (9) into three parts as 
enclosed in the dashed boxes. Evolution inside each box has remained 
shielded from the outside. The evolution in the middle box requires that 
site 3 flips up before site 2 or site 4. The probability for this event 
is equal to $\frac{1}{3}$. Given that site 2 is down, the probability 
that site 1 is down is equal to $e^{-p_0(h_a)}$. Similarly, given that 
site 4 is down the probability that site 5 is down is equal to 
$e^{-p_0(h_a)}$. Thus the probability per site of observing two adjacent 
doublets on ramp-I at an applied field $h_a$ is equal to 
$\frac{1}{3}e^{-2p_0(h_a)}$. Note that it is quite different from the 
square of the probability of finding a single doublet! Let $h_{1}$, 
$h_{2}$, $h_{3}$, $h_{4}$ and $h_{5}$ denote the quenched fields at 
sites 1, 2, 3, 4, and 5 respectively. We are interested in the case 
$h_{2}>h_{1}$, and $h_{4}>h_{5}$, and ask what are {\em{a posteriori}} 
distribution of fields $h_{1},h_{2},\ldots,h_{5}$ on ramp-II. Let the 
probability that $h_i (i=1,\ldots,5)$ lies in the range 
$[-h_a-dh_a,-h_a]$ on ramp-II be denoted by $\rho_i (-h_a) dh_a$. The 
distributions of $h_1$ and $h_5$ are given by,

\be \rho_1(-h_a)= \frac{\{1-e^{-p_1(h_a)}\}}{e^{-p_0(h_a)}} \phi(-h_a) 
\ee
\be \rho_5(-h_a)= \frac{\{1-e^{-p_1(h_a)}\}}{e^{-p_0(h_a)}} 
\phi(-h_a)\ee

Equation (21) is obtained as follows. If sites 1 and 2 are down on 
plateau-I, they must have been down all along ramp-I. Given that site 2 
is down on ramp-I, the probability that site-1 is down is equal to 
$e^{-p_0(h_a)}$. This accounts for the denominator. The numerator gives 
the {\em{a posteriori}} probability distribution of the quenched field 
at site 1 on the lines of the preceding discussion with reference to 
figure (8). Equation (22) is written similarly. Note that the 
distributions $\rho_1(h_i)$ and $\rho_5(h_i)$ are each normalized to 
unity. Next we turn to the distributions of $h_2$, $h_3$, and $h_4$. As 
stated previously, $h_{2}>h_{1}$ and $h_{4}>h_{5}$ so we are looking at 
the case where $h_2$ and $h_4$ flip up on ramp-II. We may assume without 
loss of generality that $h_4 > h_2$. Let $h_4$ flips up on ramp-II at 
applied field $h_a$ i.e. $h_4+h_a=0$. The probability for this is equal 
to $\phi(-h_a) p_1(h_a) \{1-p_1(h_a)\}$; the three multiplicative 
factors giving respectively the probability that $h_4=-h_a$, site 3 is 
up at $h_a$, and site 2 is down at $h_a$. Thus,

\be \rho_4(-h_a)= 6 p_1(h_a) \{1-p_1(h_a)\} \phi(-h_a)\ee

The factor of 6 on the rhs arises as follows. At $h_a$ on ramp-II site 2 
will be down and sites 3 and 4 will be up at $h_a$ given that at an 
earlier field $h_a-2|J|$ on ramp-I site 3 had flipped up before 2 and 4 
with probability $\frac{1}{3}$. The site flipping up on ramp-II may be 
to the left of the central site or to its right. This gives an 
additional factor of 2.

The distribution of $h_{2}$ is obtained similarly. We find,

\be \rho_{2}(-h_a)=3 p_1^{2}(h_a) \phi(-h_a) \ee

\subsection{Ramp-II}

Ramp-II is determined by the combination of two terms. The dominant term 
is the increase in magnetization due to the decrease in the number of 
doublets. When a doublet disappears, it adds an extra up spin in the 
system which increases the magnetization. Occasionally, a disappearing 
doublet creates a string of three up spins. A triplet of up spins is 
unstable on ramp-II if $\Delta\le|J|$ and therefore the central spin of 
the triplet flips down as soon as the triplet is created. This decreases 
the magnetization. In the following, we calculate the above two terms 
separately. Refer to figure (8) for calculating the first term. Let us 
assume $h_1< h_2$. The {\em{a posteriori}} distribution of fields $h_1$ 
and $h_2$ in figure (8) are the same as $\rho_1(h_1)$ in figure (9). The 
probability that the doublet disappears at $h_2+h_a=0$ on ramp-II is 
given by

\be P_{\uparrow\uparrow}^{II}=2 e^{-2p_0(h_a)} 
\int_{-h_a}^{\infty}\rho_1(h_2) dh_2 \int_{-\infty}^{h_2}\rho_1(h_1) 
dh_1 \ee

The factor $e^{-2p_0(h_a)}$ is the probability per site of finding a 
doublet at $h_a$ before any of them have been relaxed. The factor 2 
takes care of the fact that either $h_{1}$ or $h_{2}$ may be the larger 
field although the expression is written on the assumption that site 2 
flips up first. When a doublet disappears, a pair of adjacent up spins 
is created. This is the reason for the choice of the subscript on 
$P_{\uparrow\uparrow}^{II}$. The superscript indicates that the 
probability refers to ramp-II. We obtain,

\be
P_{\uparrow\uparrow}^{II}= e^{-2p_0(h_a)} \left.{\left\{ 
\int_{-\infty}^{h_2}\rho_1(h_1) dh_1\right\}}^2 
\right|^{h_2=\infty}_{h_2=-h_a} 
\ee

For $\Delta \le |J|$, $p_0(h_a)=1$ if $p_1(h_a)> 0$. Therefore,

\be P_{\uparrow\uparrow}^{II}= \frac{1}{e^2} - 
\left[\left(1+e^{-1}\right) - \left\{p_1(h_a)+e^{-p_1(h_a)}\right\} 
\right]^{2} \ee

We now calculate the fraction of (unstable) up triplets on ramp-II. Refer 
to figure (9) with the assumption that $h_2 < h_4$. An up triplet forms 
when $h_2+h_a=0$. The cumulative fraction of up triplets is 
given by,

\be P_{\uparrow\uparrow\uparrow}^{II}=\frac{1}{3} e^{-2p_0(h_a)} 
\int_{-h_a}^{\infty}\rho_2(h_2) dh_2 \int_{-\infty}^{h_2}\rho_1(h_1) 
dh_1 \int_{h_2}^{\infty}\tilde\rho_4(h_4) dh_4 
\int_{-\infty}^{h_4}\rho_5(h_5) dh_5 \ee

The two factors before the integrals give the probability per site of 
finding the object shown in Figure (9). These take into account the 
condition that $h_3>h_4$. The next two integrals give the probability 
that $h_2+h_a=0$ and $h_1 < h_2$. When $h_2$ is in the range $-h_a$ and 
$-h_a-\delta h_a$, $h_4$ can be anywhere in the range $h_2$ to $\infty$. 
Let $\tilde{\rho_4(h_4)}$ be the density of $h_4$ in this range. 
Clearly,

\be \rho_2(h_2)=\int_{h_2}^{\infty} \tilde{\rho}_4(h_4) dh_4, 
\mbox{   or   }\tilde{\rho}_4(h_i)=-\frac{d\rho_2(h_i)}{ dh_i} \ee.

The integrals in equation (28) can be evaluated exactly for a uniform 
distribution. We get a non-zero contribution only if $p_1(h_a) >0$. If 
$\Delta < |J|$ and $p_1(h_a) >0$ then we must necessarily have 
$p_0(h_a)=1$. Thus we get,

\bdm P_{\uparrow\uparrow\uparrow}^{II} =\frac{1}{3} \left[
        \frac{3}{2} + \frac{6}{e}
        - 6 \left\{ 1 + \frac{1}{e} \right\} p_1(h_a)
        + 3 p_1^{2}(h_a)
        + {\left\{ 1 + \frac{1}{e} \right\}}^{2} p_1^{3}(h_a)
        -\frac{5}{4} \left\{ 1 + \frac{1}{e} \right\} p_1^{4}(h_a)
        +\frac{2}{5} p_1^{5}(h_a) \right.  \\\edm

\bdm
         \left. -\left\{ 6 \left( 1 + \frac{1}{e} \right)
        -6 p_1(h_a) -3 \left( 1 + \frac{1}{e} \right) p_1^{2}(h_a)
        + 2 p_1^{3}(h_a) \right\} e^{-p_1(h_a)}
        +\left\{\frac{9}{2} + 3 p_1(h_a)\right\} e^{-2p_1(h_a)}
\right]
\edm

Putting the various terms together, the probability that a randomly 
chosen spin on the lattice is up on ramp-II is given by

\be
P_{\uparrow}^{II}(h_a) = \frac{1}{2}[1-e^{-2}] +
P_{\uparrow\uparrow}^{II}(h_a) -
P_{\uparrow\uparrow\uparrow}^{II}(h_a)
\ee

The magnetization on ramp-II is given by
\be
m^{II}(h_a)=2 P_{\uparrow}^{II}(h_a) -1
\ee

The magnetization on plateau-II is equal to,

\be
m^{II}= \left[ \frac{27}{30} -\frac{7}{6} e^{-1}
-\frac{8}{3} e^{-2} \right] = .109
\mbox{  (approximately) }
\ee


\subsection{Plateau-II}

Each down spin on plateau-II is a singlet. However, there are three 
different classes of singlets: the singlets formed on ramp-I; singlets 
formed on ramp-II by a vanishing doublet; and finally the singlets 
formed on ramp-II by the unstable central spin of an up triplet flipping 
down. The {\em{a posteriori}} distribution of random field is different 
for each class. Let $\rho^{II}_a$, $\rho^{II}_b$, and $\rho^{II}_c$ 
denote the density of random fields in the three cases and $P^{II}_a$, 
$P^{II}_b$, and $P^{II}_c$ be the probability of finding the 
corresponding singlet at applied field $h_a$on plateau-II.

\be P^{II}_x(h_a)=\int_{-h_a}^{\infty}\rho^{II}_x(h_i) dh_i, \mbox{ x=a, 
b, c } \ee

It is useful to think of singlets in each class as being black or white 
on the ramp on which they are created. Suppose a singlet is created on 
ramp-I at applied field $h_a^\prime$. It is created black if 
$h_i+2|J|+h_a^\prime \ge 0$, and white if $h_i+2|J|+h_a^\prime < 0$ 
where $h_{i}$ is the quenched random field at the singlet site. If a 
singlet is black at applied field $h_a^\prime$ then it is also black at 
fields greater than $h_a^\prime$. A black singlet on ramp-I at 
$h_a^\prime$ will turn up on ramp-III at $h_a=h_a^\prime+4|J|$. 
Therefore if we know the fraction of black singlets on ramp-I we can 
calculate how they are destroyed on ramp-III.

The fraction of black singlets created on ramp-I is given by,

\be 
P^{II}_a(h_a^\prime)=p_0(h_a^\prime)-\frac{1}{2}\left[1-e^{-2p_0(h_a^\prime)}\right] 
- 2 e^{-p_0(h_a^\prime)}\left[e^{-p_0(h_a^\prime)}-\{1-p_0(h_a^\prime)\}\right] 
\ee

The explanation of the above equation is as follows. Imagine ordering 
the sites of the lattice in order of decreasing quenched field on the 
site. When all sites with $h_i+2|J|+h_a^\prime \ge 0$ have been relaxed, 
the fraction of the relaxed sites is equal to $p_0(h_a^\prime)$ (the 
black sites). This fraction is made of the up sites (the second term on 
the right), black doublet sites (the last term), and the black singlets. 
Hence the equation for $P^{II}_{1}(h_a^\prime)$. The last term is 
written as follows. In each doublet, there are two sites from which we 
can choose one. This accounts for the factor 2. The quantity in the 
square bracket gives the probability that the chosen site is black, and 
$e^{-p_0(h_a^\prime)}$ is the probability that the other site can have 
any allowed value of the quenched field.

It is instructive to derive equation (34) by an alternate and more 
direct method as well. We note that a pair of down spins on the chain 
has to be followed by a down spin or an up spin, i.e. 
$P_{\downarrow\downarrow}= P_{\downarrow\downarrow\downarrow} + 
P_{\downarrow\downarrow\uparrow}$. Similarly, 
$P_{\uparrow\downarrow\downarrow} + P_{\downarrow\downarrow\downarrow} = 
P_{\downarrow\downarrow}$. Thus,

\be P_{\uparrow\downarrow\downarrow}(h_a^\prime) = 
P_{\downarrow\downarrow\uparrow}(h_a^\prime) = 
P_{\downarrow\downarrow}(h_a^\prime) - 
P_{\downarrow\downarrow\downarrow}(h_a^\prime) 
=e^{-2p_0(h_a^\prime)}-\{1-p_0(h_a^\prime)\} 
e^{-2p_0(h_a^\prime)} =p_0(h_a^\prime)e^{-2p_0(h_a^\prime)} \ee

We can obtain the fraction of singlets $P_{\uparrow\downarrow\uparrow}$ 
from $P_{\uparrow\downarrow\downarrow}$ and 
$P_{\downarrow\downarrow\uparrow}$ by calculating the probability that 
the down spin at either end flips up under a $p_0$ process.

\be P_{\uparrow\downarrow\uparrow}(h_a^\prime)= 
2\int_{-\infty}^{h_a^\prime} 
P_{\uparrow\downarrow\downarrow}(h_a^{\prime\prime}) 
\phi(-2|J|-h_a^{\prime\prime}) dh_a^{\prime\prime} = 2 
\int_0^{p_0(h_a^\prime)}e^{-2p_0}p_0 dp_0 =\frac{1}{2} 
\left[1-e^{-2p_0(h_a^\prime)}\right]-p_0(h_a^\prime)e^{-2p_0(h_a^\prime)} 
\ee

The above expression gives the total fraction of singlets on ramp-I 
which include black $(h_i+2|J|+h_a^\prime\ge0)$ as well as white 
$(h_i+2|J|+h_a^\prime<0)$ singlets. The fraction of white singlets is 
given by,

\be P_{\uparrow\downarrow\uparrow}^{white}(h_a^\prime)= 
\{1-p_0(h_a^\prime)\}[1-e^{-p_0(h_a^\prime)}]^2 \ee

The first factor on the rhs gives the probability that the central site 
is white. Given that the central site is white, the probability that its 
neighbor is up is equal to $1-e^{-p_0(h_a^\prime)}$. The probability 
that both neighbors are up is equal to $[1-e^{-p_0(h_a^\prime)}]^2$. Thus, 

\be P_{\uparrow\downarrow\uparrow}^{black}(h_a^\prime)= 
P_{\uparrow\downarrow\uparrow}(h_a^\prime) - 
P_{\uparrow\downarrow\uparrow}^{white}(h_a^\prime)=p_0(h_a^\prime)-\frac{1}{2}\left[1-e^{-2p_0(h_a^\prime)}\right] 
- 2 
e^{-p_0(h_a^\prime)}\left[e^{-p_0(h_a^\prime)}-\{1-p_0(h_a^\prime)\}\right] 
\ee

The fraction of black singlets on ramp-II ($h_i+h_a^{\prime\prime}\ge0$) 
generated by vanishing doublets is given by,

\be P^{II}_b(h_a^{\prime\prime})= 
\left[e^{-p_1(h_a^{\prime\prime})}-\{1-p_1(h_a^{\prime\prime})\}\right]^2 
\ee

The above equation is easily understood. It is the probability that both 
sites of the doublet are black. If both sites of the doublet are black, 
the one with higher random field must flip up on ramp-II leaving us 
with a singlet on plateau-II which is black.

The fraction of black singlets created by unstable triplets requires the 
calculation of triplets. We have calculated the fraction of triplets as 
they are formed on ramp-II. What we need now is a similar but different 
calculation. The point can be understood with a reference to Figure (9). 
Recall that $h_{3} \ge h_{4} \ge h_{2}$. On ramp-II, we needed the 
fraction of triplets with $h_{2}+h_a^{\prime\prime} \ge 0$, because the 
formation of triplets is controlled by this threshold. The restoration 
of the triplets on ramp-III is controlled by the condition 
$h_{3}-2|J|+h_a \ge 0$. Keeping in mind that we want $h_{1} \le h_{2}$, 
and $h_{5} \le h_{4}$, the probability that $h_3-2|J|+h_a \ge 0$ is 
given by,

\bdm
P^{II}_c(h_a)=2 \int_{-h_a+2|J|}^{\infty}\phi(h_3)dh_3
\int_{-\infty}^{h_3}\rho_4(h_4) dh_4
\int_{-\infty}^{h_4} \rho_5(h_5) dh_5
\int_{-\infty}^{h_4} \rho_2(h_2) dh_2
\int_{-\infty}^{h_2} \rho_1(h_1) dh_1
\edm

Using $p_0(h_a)=1$ and $p_1(h_a)=1$ on plateau-II we get,

\bea
P^{II}_c(h_a)= & -\left( 1 + \frac{2}{e} \right)
+ \left[ \frac{1}{4} + \frac{2}{e}
+ \frac{4}{e^{2}} \right] p_2(h_a)
- \frac{1}{2} \left[ 1 + \frac{5}{e}
+ \frac{4}{e^{2}} \right] p_2^2(h_a)
+ \frac{1}{3} \left[ \frac{3}{2}
+ \frac{4}{e} + \frac{1}{e^{2}}
\right] p_2^3(h_a) \nonumber \\ &
-\frac{1}{4} \left[ 1+\frac{1}{e} \right] p_2^4(h_a)
+ \frac{1}{20} p_2^5(h_a) + 
\left[1+\frac{2}{e}\right]e^{-p_2(h_a)}
-\frac{2}{e}p_2(h_a) e^{-p_2(h_a)} +p_2^2(h_a) e^{-p_2(h_a)}\nonumber \\ 
&
+\frac{1}{2} \left[ 1-e^{-2p_2(h_a)} \right]
\eea

\subsection{Ramp-III}

The rise of magnetization on ramp-III is due to singlet sites turning up 
in increasing field. At the start of ramp-III there are three categories 
of singlets present on plateau-II, and we have classified each of them 
conveniently into black and white singlets. The fraction of singlets on 
plateau-II that turn up at $h_a$ on ramp-III is given by the fraction of 
black singlets in each of the three categories:

\be P^{III}(h_a)=P^{III}_a(h_a)+P^{III}_b(h_a)+P^{III}_c(h_a)\ee

However the calculation of magnetization on ramp-III turns out to be a 
bit more complicated. There is a new twist. Frequently when an original 
singlet site on plateau-II turns up on ramp-III its nearest neighbor 
turns down. We call this the creation of a new singlet on ramp-III. The 
newly created singlet site would turn up at a larger applied field on 
ramp-III. We shall call this event the destruction of the newly created 
singlet. We have to calculate the newly created singlets and their 
destruction before the magnetization on ramp-III may be obtained. It 
should be noted that this means that some sites flip three times in the 
course of a monotonic increase of applied field from $-\infty$ to 
$\infty$. However, no site flips more than three times.

When does a vanishing singlet on ramp-III create a new singlet on an 
adjacent site? Consider the singlet at site 3 in figure 10. Suppose site 
3 flips up at $h_{a}$, i.e. $h_3-2|J|+h_a = 0$. Now the net field on 
site 2 is equal to $h_2 + h_a$. This is necessarily positive because 
$h_2+2|J|-h_3\ge0$ if $\Delta \le |J|$. Thus site 2 would stay up after 
site 3 flips up. Consider site 4. After site 3 has turned up the net 
field at site 4 is equal to $h_4-2|J| +h_a$. Thus site 4 will turn down 
if $h_4<h_3$. Site 3 will stay up even if site 4 turns down because 
$h_3+h_a>0$. These considerations can be put in the form of two guiding 
rules. When a singlet turns up on ramp-III, (i) its nearest neighbor 
stays up if the next nearest neighbor is down, and $\Delta \le |J|$, 
(ii) its nearest neighbor turns down if it has less quenched field than 
the singlet and the next nearest neighbor is up. Detailed considerations 
show that only the singlets created on ramp-I fall under the purview of 
these rules. Therefore we focus on the singlets present on plateau-I. 
Specifically we focus on the configurations shown in figure (11) and 
figure (12). In each of these figures a new singlet is created on site 3 
when the singlet on site 2 is destroyed. The two figures make different 
contributions because of the role played by the next nearest neighbor of 
the singlet on the side of the newly created singlet.

The {\em{a posteriori}} distribution of the quenched field at site 2 is 
given by \be 
\tilde{\tilde\phi}(h_2)=\left[1-e^{-p_2(-h_2+2|J|)}\right]\phi(h_2)\ee

The contributions of figure (11) to the fraction of newly created 
singlets when all sites with $h_i-2|J|+h_a \ge 0$ have been relaxed on 
ramp-III is given by,

\bea P^{III}_d(h_a) = 2 
\int_{-h_a+2|J|}^{\infty}\tilde{\tilde\phi}(h_2)dh_2 
\int_{-\infty}^{h_2}\phi(h_3)dh_3 
\int_{h_2}^{\infty}\tilde{\tilde\phi}(h_4)dh_4 \nonumber \\ = 
\frac{3}{2} -2p_2(1-p_2) -\frac{2}{3} p_2^{3} -2 (1-p_2+p_2^2) e^{-p_2} 
+\frac{1}{2}(1-2p_2) e^{-2p_2}\mbox{     } [p_2\equiv p_2(h_a)]. 
\nonumber 
\\ \eea

Similarly the contribution of figure (12) is, 

\bea P^{III}_e(h_a) =  
\int_{-h_a+2|J|}^{\infty}\tilde{\tilde\phi}(h_2)dh_2 
\int_{-\infty}^{h_2}\phi(h_3)dh_3
\int_{-\infty}^{h_3}\phi(h_4)dh4 
\int_{-\infty}^{h_4}\tilde{\tilde\phi}(h_5)dh_5 \nonumber \\  
=-\left(\frac{1}{3} + 3 e^{-1} \right)
+\left(\frac{1}{3} + 5 e^{-1} \right) p_2
-\left(\frac{1}{2} + 2 e^{-1} \right) p^{2}_2
+\frac{1}{3}\left( 1 + e^{-1} \right) p^{3}_2 \nonumber \\ 
-\frac{1}{12} p^{4}_2 
+\left\{ \left(\frac{4}{3} + 3 e^{-1} \right)
-\left( 1 + 2 e^{-1} \right) p_2 + e^{-1} p^{2}_2
-\frac{1}{3} p^{3}_2 \right \} e^{-p_2} - e^{-2p_2} \nonumber \\
\eea

In calculating the total fraction of newly created singlets, we have to 
multiply the contribution of figure (12) by 2 because an equal 
contribution is made by a configuration in which the doublet is to the 
left of the vanishing singlet.

The destruction of newly created singlets on ramp-III can be analyzed in a
similar manner as their creation. The probability that site 3 flips for 
the third time in figure (11) is given by,

\bea P^{III}_f(h_a) = 2 \int_{-h_a+2|J|}^{\infty}\phi(h_3)dh_3 
\int_{h_3}^{\infty}\tilde{\tilde\phi}(h_2)dh_2 
\int_{h_3}^{\infty}\tilde{\tilde\phi}(h_4)dh_4 \nonumber \\ =\frac{1}{2} 
\left[ 1 -e^{-2p_2} \right] -2 p_2 e^{-p_2} + p_2 (1-p_2) + \frac{1}{3} 
p^{3}_2 \nonumber \\ \eea

Similarly, the contribution to destruction of newly created singlet in 
figure (12) and another similar figure in which the doublet is to the 
left of the singlet is given by,

\bea P^{III}_g(h_a) =2  
\int_{-h_a+2|J|}^{\infty}\phi(h_3)dh_3 
\int_{h_3}^{\infty}\tilde{\tilde\phi}(h_2)dh_2
\int_{-\infty}^{h_3}\phi(h_4)dh4 
\int_{-\infty}^{h_4}\tilde{\tilde\phi}(h_5)dh_5 \nonumber \\  
= 2 e^{-1} - \left(1 + 4 e^{-1} \right) p_2
+\left(\frac{3}{2} + 3 e^{-1} \right) p^{2}_2
-\left( 1  + \frac{2}{3} e^{-1} \right) p^{3}_2 +\frac{1}{4}p^{4}_2
\nonumber \\ 
- \left\{ \left( 1 + 2 e^{-1} \right)
- 2 \left( 1 +  e^{-1} \right) p_2 + p^{2}_2 \right \} e^{-p_2} + 
e^{-2p_2}
\nonumber \\ 
\eea

Putting all the terms together, the probability that a randomly chosen 
site on ramp-III is up is given by,

\be P^{III}_\uparrow(h_a)=P^{II}_{\uparrow}(h_a) 
+P^{III}_a(h_a)+P^{III}_b(h_a)+P^{III}_c(h_a) 
-P^{III}_d(h_a)-P^{III}_e(h_a)+P^{III}_f(h_a)+P^{III}_g(h_a) \ee

The magnetization in increasing field $h_a$ is given by $m(h_a)=2 
P_{\uparrow}^{III}(h_a) -1$. The magnetization $m_R$ on the return 
trajectory in decreasing field may be obtained by symmetry 
$m_R(h_a)=-m(-h_a)$. These results has been superimposed on the 
corresponding simulation data in figure (1) and figure (2). The 
simulation results for the return hysteresis loop were obtained 
independently without using the symmetry.

\newpage

\section{Unbounded distribution of quenched field}

It is convenient to introduce the following nomenclature. We say that a 
site flips up under a $p_0$-process if none of its two nearest neighbors 
are up when it flips up. It is said to flip up under a $p_1$-process if 
it flips up when one of its neighbors is up and the other is down. 
Similarly, a site flipping up under a $p_2$-process has both its 
neighbors up at the time it flips up. The simplifying feature of the 
analysis for $\Delta \le |J|$ is that a $p_n$-process ($n=1,2$) can not 
take place anywhere on the chain unless all $p_{n-1}$ processes have 
been exhausted. In other words a new ramp can not begin before the 
previous ramp is completed. This feature is lost if $\Delta > |J|$ (or 
if the distribution is Gaussian) because a $p_1$ or a $p_2$ process can 
occur on the chain even if $p_0$-processes have not been exhausted (i.e. 
there remain strings of three down spins on the chain). The fact that 
$p_0$, $p_1$, and $p_2$ processes can run concurrently makes the 
calculation of {\em{a posteriori}} distribution of quenched fields at 
down sites more complicated than encountered in the preceding section.

Our first task is to calculate the probability of occurrence of a 
doublet in the chain. A doublet is a pair of adjacent down spins that 
have remained down in a monotonically increasing field from $-\infty$ to 
$h_a$. The significance of a doublet lies in its screening property. It 
separates the chain into two parts that have evolved independently of 
each other. The doublet also provides a natural length in the analysis 
of the chain. This is not a fixed length but rather a variable length of 
a segment of chain that is free of doublets and lies between two 
doublets at an applied field $h_a$. The history of evolution of this 
segment may be analyzed independently of the rest of the chain. Let us 
focus on one end of such a segment. We label the sites $0, 1, 2, 3, 
\ldots, n$ starting from the left end of the segment. Our immediate 
object is to calculate the probability (per site of the chain) that 
site-0 and site-1 form a doublet at $h_a$. This is equal to 
$P_{\downarrow \downarrow}^2(1\downarrow|0\downarrow; h_a)$ where 
$P_{\downarrow\downarrow} (1\downarrow|0\downarrow; h_a)$ is the 
conditional probability that site-1 is down given that site-0 is down. 
We proved in section IV that $P_{\downarrow 
\downarrow}(1\downarrow|0\downarrow; h_a) = e^{-p_0(h_a)}$ if only 
$p_0$-processes are allowed. This is an exact result on ramp-I for 
$\Delta \le |J|$ and we take it as the leading term of the exact result 
for $\Delta > |J|$ or for an unbounded distribution such as the Gaussian 
distribution. The main effect of $\Delta<|J|$ is that if one of two 
adjacent sites say site-1 and site-2 ($h_1+2|J|+h_a \ge 0$ and 
$h_2+2|J|+h_a \ge 0$) flips up under a $p_0$-process then it prevents 
the other from doing the same. As long as $|h_1-h_2|<2|J|$ the same 
effect will be obtained at site-1 and site-2 for other distributions of 
the quenched field. In any particular realization of the quenched field 
distribution, the occurrence of a large connected cluster of sites with 
$h_{i+1}-h_i \ge 2|J|$ is rare. Therefore the case $\Delta<|J|$ serves 
as a good starting point for the exact result. Thus,

\be P_{\downarrow \downarrow}(1\downarrow|0\downarrow:h_a) = 
e^{-p_0(h_a)} \hspace{2cm}\{ \mbox{ leading 
term } \} \ee

Our approach is to add correction terms to the leading term to make it 
an exact result. The correction terms are functions of $p_1(h_a)$. No 
correction to the leading term is required if $p_1(h_a)=0$. For 
$p_1(h_a)>0$, the corrections may be divided into two categories. The 
first category is the one in which a doublet created by a $p_0$-process 
is subsequently destroyed by a $p_1$-process. This is similar to the 
fate of doublets on ramp-II for $\Delta \le |J|$ and we get,

\be P_{\downarrow \downarrow}(1\downarrow|0\downarrow; h_a) = 
e^{-p_0(h_a)}-\left[e^{-p_1(h_a)}-\left\{1-p_1(h_a)\right\}\right] 
\hspace{1cm}\{\mbox{ first order correction } \} \ee

The second category of correction involves events where a $p_1$-process 
pre-empts a $p_0$-process. These events are a signature of $\Delta >
|J|$ and do not exist if $\Delta \le |J|$. Such events are always 
possible if the quenched field has a Gaussian distribution. We illustrate this by a
simple example. Suppose for a particular realization of the distribution 
of quenched fields $h_1,h_2,h_3,h_4\ldots$, site-1, site-2, site-3 are 
black, site-4 is white, and $h_3>h_2>h_1$. If $\Delta < |J|$, site-3 
will flip up first, block site-2 from flipping next, and therefore 
site-1 will flip up last. This will result in the absence of doublet at 
site-0 and site-1. Now consider $\Delta > |J|$ so that it is possible to 
have $h_2> h_1+2|J|$ but $h_1+h_a < 0$. In this scenario site-3 will 
flip up first, then site-2, and site-1 will remain down giving us a 
doublet at site-0 and site-1 that was not allowed if $\Delta < |J|$. The 
probability of this new doublet created by a $p_1$-process pre-emptying 
a $p_0$-process is given by,

\be T_3(h_a)= 
\int_{-2|J|-h_a}^{-h_a}\phi(h_1)dh_1\int_{h_1+2|J|}^{\infty} 
\tilde\phi(h_2)dh_2 
\ee
 
Here $\tilde\phi(h_2)$ is given by equation (19) of section IV. This is 
because equation (19) of section IV gives the {\em{a posteriori}} 
distribution of the quenched field on a black site next to a site that 
has flipped up by a $p_0$-process. In our example site-2 is a black site 
next to site-3 that has flipped up by a $p_0$-process. $T_3(h_a)$ term 
is the leading term in the category of correction terms that arise 
because a $p_1$-process has pre-empted a $p_0$-process.

\be P_{\downarrow \downarrow}(1\downarrow|0\downarrow; h_a) = 
e^{-p_0(h_a)}-\left[e^{-p_1(h_a)}-\left\{1-p_1(h_a)\right\}\right] 
+T_3(h_a)\hspace{1cm}\{\mbox{ second order correction } \}\ee

The next correction comes from a cluster of adjacent spins that flip up 
as follows. Site-4 flips up under a $p_0$-process ($h_4+2|J|+h_a\ge0$), 
site-3 flips next under a $p_1$-process ($h_3>h_2+2|J|$), site-2 also 
flips under a $p_1$-process ($h_2>h_1+2|J|$) after site-3 has flipped 
up, and site-1 remains down ($h_1+h_a<0,h_1+2|J|+h_a\ge0$). The 
contribution from this event is,

\be T_4(h_a)= -  
\int_{-2|J|-h_a}^{-h_a}\phi(h_1)dh_1 
\int_{h_1+2|J|}^{\infty}\phi(h_2)dh_2 
\int_{h_2+2|J|}^{\infty}\tilde\phi(h_3)dh_3 
\ee

Notice that $T_3(h_a)$ is positive and $T_4(h_a)$ is negative. 
$T_3(h_a)$ is positive because it produces a doublet (at site-0 and 
site-1) where it did not exist under the $p_0$-process alone. $T_4(h_a)$ 
is negative for the following reason. $T_4$-process and $p_0$-process is 
mutually exclusive and since both contribute to a doublet then they must 
be added separately avoiding double counting. Thus the leading term in 
equation (48) under $p_0$-process alone is an overestimate and has to be 
reduced by an amount $T_4(h_a)$. However, $T_4$-process also gives rise 
to a doublet. Hence an amount equal to the one subtracted from the 
leading term has to be added to it resulting in zero correction to the 
leading term. Now consider the first correction shown in equation (49). 
This is an underestimate because it comes from the destruction of 
overestimated doublets in the leading term. It can be corrected by 
adding $T_4(h_a)$. The correct result at the present level of accuracy 
is,

\be P_{\downarrow \downarrow}(1\downarrow|0\downarrow; h_a) = 
e^{-p_0(h_a)}-\left[e^{-p_1(h_a)}-\left\{1-p_1(h_a)\right\}\right] 
+T_3(h_a)-T_4(h_a)\hspace{1cm}\{\mbox{ third order correction } \}\ee

Continuing in this vein we get the following exact result,

\be P_{\downarrow \downarrow}(1\downarrow|0\downarrow; h_a) = 
e^{-p_0(h_a)}-\left[e^{-p_1(h_a)}-\left\{1-p_1(h_a)\right\}\right] 
-\sum_{n=3}^{\infty}(-1)^nT_n(h_a)\hspace{1cm}\{\mbox{ exact result } \}\ee

where
\be
T_n(h_a)=\int_{-2|J|-h_a}^{-h_a}
\phi(h_1)dh_1\left[\prod_{m=3}^{n-1} 
\int_{h_{m-2}+2|J|}^{\infty}\phi(h_{m-1})dh_{m-1}\right] 
\int_{h_{n-2}+2|J|}^{\infty}\tilde\phi(h_{n-1})dh_{n-1}
\ee

The probability per site of finding a doublet on the chain is equal to,

\be P_{\downarrow\downarrow}(h_a)= \left[P_{\downarrow 
\downarrow}(1\downarrow|0\downarrow:h_a)\right]^2 \ee 

Note that although the (unconditional) probability of a doublet is the 
product of two mutually conditional probabilities, our notation for this 
is $P_{\downarrow\downarrow}(h_a)$ (without the square sign). In general 
it is not possible to do the integrals exactly to get an analytic 
expressions for $T_n(h_a)$ in a closed form. These have to be evaluated 
numerically. However $T_n$ decreases exponentially with increasing $n$. 
In our numerical work we included terms up to $n \le 4$ in the 
calculation of the conditional probability $P_{\downarrow 
\downarrow}(1\downarrow|0\downarrow:h_a)$, and used the square of this 
quantity to calculate the probability per site of doublets. This gives 
an excellent fit with the simulation data as shown in section VI.

Although a doublet at an applied field $h_a$ is the most basic object in 
our calculations but we need to calculate the probability of several 
other objects before we can calculate the magnetization curve. In order 
to calculate the magnetization curve we need to know the probability 
$P_{\uparrow}(h_a)$ that a randomly chosen site is up at $h_a$. On the 
lower half of the hysteresis loop, it is more convenient to focus on the 
complementary probability [1-$P_{\uparrow}(h_a)$] that a randomly chosen 
site is down. The randomly chosen site can have both its neighbors down, 
or both of them up, or one up and one down. If both neighbors of a down 
site are down, the site in question is necessarily a white site. The 
probability (per site) that a site is white is equal to $1-p_0(h_a)$. 
Given a white site, the probability that both its neighbors are down is 
given by,

\be 
P_{\downarrow\downarrow\downarrow}(h_a)=
[1-p_0(h_a)]P{\downarrow\downarrow}(h_a) 
\ee

Next, consider a down spin with one neighbor up and one down. In both 
cases there is a doublet, and the up spin is either to the left of the 
doublet or to its right. We have calculated the probability of a 
doublet as well as a triplet. From these we can obtain the 
probability of a doublet followed by an up spin using the 
equation

\be 
P_{\downarrow\downarrow\uparrow}(h_a)=
P_{\downarrow\downarrow}(h_a)-P_{\downarrow\downarrow\downarrow}(h_a) 
\ee 

Naturally $P_{\downarrow\downarrow\uparrow}(h_a) 
=P_{\uparrow\downarrow\downarrow}(h_a)$ by symmetry.

Substituting from equations (57), we get
\be 
P_{\downarrow\downarrow\uparrow}(h_a)=P_{\uparrow\downarrow\downarrow}(h_a) 
=p_0(h_a)P_{\downarrow\downarrow}(h_a) 
\ee 

Note that the up spin in the above object could have flipped up under a 
$p_0$ or a $p_1$ process. We have checked the above equations against 
numerical simulations shown in section VIII.

\section{Singlets} 

Now we calculate the probability $P_{\uparrow\downarrow\uparrow}(h_a)$ 
that a randomly chosen site is down and both its neighbors are up. Let 
the three consecutive sites that form this singlet be labeled $1, 2,3$ 
and $h_1,h_2$, and $h_3$ denote the respective quenched fields. Without 
loss of generality we can assume that just before this singlet is 
created site-1 was up and sites 2 and 3 were down as shown in figure 
(14), i.e. site-1 flips up before site-3. We have to know if site-1 has 
flipped up under a $p_0$-process or a $p_1$-process. We consider both 
these possibilities for site-1 as well as site-3. The case when site-2 
has never flipped in the course of applied field changing from $-\infty$ 
to $h_a$ is somewhat simpler to analyze. Even in this case the 
probability $P_{\uparrow\downarrow\uparrow}(h_a)$ depends on whether (i) 
both neighbors flipped up under a $p_0$-process, or (ii)both neighbors 
flipped up under a $p_1$-process, or (iii) one neighbor flipped up under 
$p_0$-process and the other under a $p_1$-process. In the case (iii) it 
is also important whether the neighbor that flipped up first flipped 
under a $p_0$-process or a $p_1$-process. Indeed we require the 
following objects before we can calculate $P_{\uparrow 
\downarrow\uparrow}(h_a)$ in the simpler case mentioned above.

\begin{itemize}

\item $P_{\uparrow\downarrow\uparrow A}$ The fraction of singlets at 
applied field $h_a$ when site-1 is up by a $p_0$ or a $p_1$-process and 
site-3 flips up by a $p_0$-process.

\item $P_{\uparrow \downarrow\uparrow B}$ Fraction of $P_{\uparrow 
\downarrow\uparrow A}$ that are black at creation.

\item $P_{\uparrow \downarrow\uparrow C}$ Fraction of $P_{\uparrow 
\downarrow\uparrow A}$ that are white at creation. 

\item $P_{\uparrow \downarrow\uparrow D}$ The fraction of singlets at 
applied field $h_a$ when site-1 is up by a $p_0$ or a $p_1$-process and 
site-3 flips up next by a $p_1$-process.

\item $P_{\uparrow \downarrow\uparrow E}$ Fraction of $P_{\uparrow 
\downarrow\uparrow D}$ that are black at creation.

\item $P_{\uparrow \downarrow\uparrow F}$ Fraction of $P_{\uparrow 
\downarrow\uparrow D}$ that are white at creation ($P_{\uparrow 
\downarrow\uparrow D}-P_{\uparrow \downarrow\uparrow E}$).

\item $P_{\uparrow\downarrow\uparrow G}$ The fraction of singlets at 
applied field $h_a$ when site-1 and site-3 flip up by a $p_0$-process and 
site-2 is white.

\item $P_{\uparrow\downarrow\uparrow H}$ The fraction of singlets at 
applied field $h_a$ when site-1 flips up by a $p_1$-process, site-3 by a 
$p_0$-process, and site-2 is white. Site-1 flips up before site-3.

\item $P_{\uparrow\downarrow\uparrow I}$ The fraction of singlets 
$P_{\uparrow\downarrow\uparrow C}$ that are white at their creation but 
are black at $h_a$.

\item $P_{\uparrow\downarrow\uparrow J}$ The fraction of singlets at 
applied field $h_a$ when site-1 and site-3 flip up by a $p_1$-process 
and site-2 is white.

\item $P_{\uparrow\downarrow\uparrow K}$ The fraction of singlets at 
applied field $h_a$ when site-1 flips up by a $p_1$-process, site-3 by a 
$p_0$-process, and site-2 is white. Site-3 flips up before site-1.

\item $P_{\uparrow\downarrow\uparrow L}$ The fraction of singlets 
$P_{\uparrow\downarrow\uparrow F}$ that are white at their creation but 
are black at $h_a$.

\item $P_{\uparrow\uparrow\uparrow M}$ Destruction of singlets 
associated with $P_{\uparrow\downarrow\uparrow B}$. By destruction we 
mean the disappearance of a singlet due to the down spin flipping up 
under a $p_2$-process.
 
\item $P_{\uparrow\uparrow\uparrow N}$ Destruction of singlets 
associated with $P_{\uparrow\downarrow\uparrow E}$. 
 
\item $P_{\uparrow\uparrow\uparrow O}$ Destruction of singlets 
associated with $P_{\uparrow\downarrow\uparrow C}$.

\item $P_{\uparrow\uparrow\uparrow P}$ Destruction of singlets 
associated with $P_{\uparrow\downarrow\uparrow F}$. 

\end{itemize}

Up to this point we have focused on singlet sites that have never 
flipped starting from the saturated state at $h_a=-\infty$. In other 
words we have considered configurations comprising site-1 up, site-2 
down, site-3 up where site-2 has never flipped up in increasing field 
from $-\infty$ to $h_a$. However, the anti-ferromagnetic dynamics also 
allows a singlet with site-2 having flipped twice i.e. site-2 can flip 
up and flip down again in the course of monotonically increasing applied 
field. In order to take into account the analysis of this second 
category of singlets, we need to calculate the following objects that 
are defined with respect to figures (15), (16), and (17).

\begin{itemize}

\item $P_{\uparrow\downarrow\uparrow Q}$ This object refers to figure 
(15) with the proviso that $h_2>h_1$, $h_4>h_5$, $h_4>h_2$ ($h_4<h_2$ 
will give an equal contribution). Site-3 flips first, site-4 flips next, 
site-2 flips after site-4 causing site-3 to flip down. 
$P_{\uparrow\downarrow \uparrow Q}$ refers to the fraction of singlets 
created in this way.

\item $P_{\uparrow\downarrow\uparrow R}$ This object refers to figure 
(16) with the proviso that $h_2>h_3$, $h_4>h_3$, $h_4>h_2$ ($h_4<h_2$ 
will give an equal contribution). Site-3 flips first, site-4 flips next, 
site-2 flips after site-4 causing site-3 to flip down. 
$P_{\uparrow\downarrow \uparrow R}$ refers to the fraction of singlets 
created in this way.

\item $P_{\uparrow\downarrow\uparrow S}$ This object refers to figure 
(17) with the proviso that $h_2<h_3$, $h_4>h_3$. Site-3 flips first, 
site-2 flips next, site-4 flips after site-2 causing site-3 to flip 
down. $P_{\uparrow\downarrow \uparrow S}$ refers to the fraction of 
singlets created in this way.

\item $P_{\uparrow\downarrow\uparrow T}$ This object also refers to 
figure (17) with the proviso that $h_2<h_3$, $h_4>h_3$. Site-3 flips 
first, site-4 flips next, site-2 flips after site-4 causing site-3 to 
flip down. $P_{\uparrow\downarrow \uparrow T}$ refers to the fraction of 
singlets created in this way.

\item $P_{\uparrow\uparrow\uparrow U}$ This object refers to the 
destruction of singlets associated with $P_{\uparrow\downarrow\uparrow 
Q}$.

\item $P_{\uparrow\uparrow\uparrow V}$ This object refers to the 
destruction of singlets associated with $P_{\uparrow\downarrow\uparrow 
R}$.

\item $P_{\uparrow\uparrow\uparrow W}$ This object refers to the 
destruction of singlets associated with $P_{\uparrow\downarrow\uparrow 
S}$.

\item $P_{\uparrow\uparrow\uparrow X}$ This object refers to the 
destruction of singlets associated with $P_{\uparrow\downarrow\uparrow 
T}$.

\end{itemize}

\subsection{$P_{\uparrow\downarrow\uparrow A}$}

In an increasing applied field the objects associated with 
$P_{\uparrow\downarrow\uparrow A}$ are created from objects associated 
with $P_{\downarrow\downarrow\uparrow}$ and $P_{\uparrow\downarrow 
\downarrow}$ when the down site at one end of these objects flips up 
under a $p_0$-process. Suppose site-3 flips up under a $p_0$-process at 
$h^\prime$ ($-\infty \le h^\prime \le h_a$). Then we have 
$h_3+2|J|+h^\prime=0$ with probability $\phi(-2|J|-h^\prime)$. Thus,

\be P_{\uparrow\downarrow\uparrow A}(h_a)= \int_{-\infty}^{h_a} 
[P_{\downarrow\downarrow\uparrow}(h')+ 
P_{\uparrow\downarrow\downarrow}(h')]\phi(-h^\prime-2|J|)dh' = 
\int_{-\infty}^{h_a} 2 
p_0(h^\prime)P_{\downarrow\downarrow}(h')\phi(-h^\prime-2|J|) 
dh^{\prime} \ee

where we have used equation (59). As a check we note that equation (59) 
is recovered by differentiating equation (60) with respect to $h_a$.

\subsection{$P_{\uparrow\downarrow\uparrow B}$}

This quantity is given by the equation.

\be 
P_{\uparrow\downarrow\uparrow B}(h_a)=\left.2\int_{-\infty}^{h_a} 
\left[ P_{\downarrow\downarrow} 
{(2\downarrow|3\downarrow:h')}-\{1-p_0(h')\}\right] 
P_{\downarrow\downarrow}{(4\downarrow|3\downarrow:h')} 
\phi(-h^\prime-2|J|) 
dh^{\prime} \right.\ee

The explanation of the above equation is as follows. The integrand 
comprises three factors. The last factor is the prob that site-3 flips 
up at $h^\prime$ under a $p_0$-process. The other factors take into 
account that site-2 and site-4 (the right neighbor of site-3) are down 
at $h^\prime$ and site-2 is black: $\phi(-h^\prime-2|J|) dh^{\prime}$ is 
the probability that site-3 flips up at $h^\prime$; 
$P_{\downarrow\downarrow} (2\downarrow|3\downarrow:h')-\{1-p_0(h')\}$ is 
the probability that site-2 is down and black ($h_2+2|J|+h^\prime > 0$); 
$P_{\downarrow\downarrow}(4\downarrow |3\downarrow; h')$ is the 
probability that site-4 is down.

\subsection{$P_{\uparrow\downarrow\uparrow C}$}

This is simply equal to $P_{\uparrow \downarrow\uparrow A}-P_{\uparrow 
\downarrow\uparrow B}$ but the notation is useful in the following 
analysis.

\be P_{\uparrow\downarrow\uparrow C}(h_a)= P_{\uparrow\downarrow\uparrow 
A}(h_a)- P_{\uparrow\downarrow\uparrow B}(h_a) \ee

\subsection{$P_{\uparrow\downarrow\uparrow D}$}

The singlets in this category are generated by doublets that are 
bordered by up spins at both ends. Thus site-1 is up, site-2 and site-3 
are down, and site-4 is up. Site-2 and site-3 are on equal footing and 
therefore the {\em{a posteriori}} distribution of quenched fields 
$\tilde\phi(h_2)$ and $\tilde\phi(h_3)$ are identical. We can assume 
without loss of generality that $h_3>h_2$ and multiply the result by a 
factor 2. Thus we focus on a singlet created at site-2 when site-3 flips 
up at $h^{\prime}$ ($-\infty < h^\prime \le h_a$). We get,

\be P_{\uparrow\downarrow\uparrow D}(h_a)=2\int_{-\infty}^{h_a} 
dh^\prime \tilde\phi(-h^\prime) p_0(h^\prime) 
P_{\downarrow\downarrow}(2\downarrow|3\downarrow; h^\prime) \ee

The explanation of the above equation is as follows: 
$\tilde\phi(-h^\prime)$ is the probability that site-3 flips up at 
$h^\prime+h_3=0$; $p_0(h^\prime) 
P_{\downarrow\downarrow}(2\downarrow|3\downarrow; h^\prime)$ is the 
probability that site-2 is down and site-1 is up just before site-3 
flips up. The {\em{ a posteriori}} distribution $\tilde\phi(h^\prime)$ 
may be obtained by differentiating the following equation.

\be \int_{-h^\prime}^{\infty} dh_3 \tilde\phi(h_3) = 
\left[e^{-p_1(h^\prime)}-\left\{1-p_1(h^\prime)\right\}\right] 
-\sum_{n=3}^{\infty}(-1)^n \tilde{T}_n(h^\prime-2|J|)\ee

where,

\be \tilde{T}_n(h^\prime)=\int_{-2|J|-h^\prime}^{\infty} 
\phi(h_1)dh_1\left[\prod_{m=3}^{n-1} 
\int_{h_{m-2}+2|J|}^{\infty}\phi(h_{m-1})dh_{m-1}\right] 
\int_{h_{n-2}+2|J|}^{\infty}\tilde\phi(h_{n-1})dh_{n-1} \ee

We may rewrite equation (54) as 

\be P_{\downarrow \downarrow}(1\downarrow|0\downarrow; h^\prime) = 
e^{-p_0(h^\prime)}-\left[e^{-p_1(h^\prime)}-\{1-p_1(h^\prime)\}\right] 
-\sum_{n=3}^{\infty}(-1)^n \left[ \tilde{T}_n(h^\prime) 
- \tilde T_n(h^\prime-2|J|)\right]\ee

For spins flipping up under a $p_1$-process, only the second and the 
last term on the right hand side come into play.

\subsection{$P_{\uparrow\downarrow\uparrow E}$}

Here we want the fraction of singlets in section D that are created 
black. If $h'$ is the field at which site-3 flips up by a $p_1$-process 
then the cumulative fraction of singlets which are black at creation is 
given by

\be P_{\uparrow\downarrow\uparrow E}(h_a)=2 \int_{-\infty}^{h_a} 
dh^\prime \tilde\phi(-h^\prime) \left[ 
P_{\downarrow\downarrow}(2\downarrow|3\downarrow; h')- 
\left\{1-p_0(h')\right\}\right] \ee

The first factor in the integrand gives the probability that 
$h_3+h^\prime=0$, and the second factor gives the probability that 
site-2 is down and black at $h^\prime$.

\subsection{$P_{\uparrow\downarrow\uparrow F}$}

\be P_{\uparrow\downarrow\uparrow F}(h_a)= P_{\uparrow\downarrow\uparrow 
D}(h_a)-P_{\uparrow\downarrow\uparrow E}(h_a) \ee

\subsection{$P_{\uparrow\downarrow\uparrow G}$}

If a site is white with probability $1-p_0(h_a)$ then it is down with 
probability unity. The probability that either of its neighbor is down 
under a $p_0$-process alone is equal to $e^{-p_0(h_a)}$. Thus the 
probability that either of its neighbor is up under a $p_0$-process 
alone is equal to $1-e^{-p_0(ha)}$. Therefore,

\be
P_{\uparrow\downarrow\uparrow G}(h_a)=\{1-p_0(h_a)\}[1-e^{-p_0(h_a)}]^2
\ee

\subsection{$P_{\uparrow\downarrow\uparrow H}$}

\be P_{\uparrow\downarrow\uparrow H}(h_a)=2 
\{1-p_0(h_a)\}\int_{-\infty}^{h_a}dh^\prime 
P_{\downarrow\downarrow}(3\downarrow|2\downarrow; h^\prime) 
\phi(-2|J|-h^{\prime}) \int_{-\infty}^{h^\prime} 
\tilde\phi(-h^{\prime\prime}) dh^{\prime\prime} \ee

The above equation is understood as follows. The integral over 
$h^{\prime\prime}$ takes care of the site flipping up under a 
$p_1$-process at $h^{\prime\prime}$. Therefore the density associated 
with this integral is the {\em{a posteriori}} distribution 
$\tilde\phi(-h^{\prime\prime})$. Site flipping up under the 
$p_0$-process flips up at $h^{\prime}$ where $h^\prime > 
h^{\prime\prime}$.The density associated with this site is 
$P_{\downarrow\downarrow}(3\downarrow|2\downarrow; h^\prime) 
\phi(-2|J|-h^{\prime})$. The first factor accounts for the fact that the 
site in question is down given that it is next to a down site (the white 
site with probability $1-p_0(h_a)$ and the second factor accounts for 
the fact that although it is down it is on the verge of turning up at 
$h^\prime$ under a $p_0$-process. Finally the factor 2 takes care of an 
equivalent configuration in which the locations of sites flipping up 
under a $p_0$ and a $p_1$ process are interchanged.

\subsection{$P_{\uparrow\downarrow\uparrow I}$}

The fraction of singlets $P_{\uparrow\downarrow\uparrow C}$ that are 
white at their creation but black at $h_a$ is given by,

\be P_{\uparrow\downarrow\uparrow I}= P_{\uparrow\downarrow\uparrow C} - 
P_{\uparrow\downarrow\uparrow G} -  P_{\uparrow\downarrow\uparrow H} \ee

\subsection{$P_{\uparrow\downarrow\uparrow J}$}

\be P_{\uparrow\downarrow\uparrow J}(h_a)= [1-p_0(h_a)] \left[ 
\int_{-h_a}^{\infty}dh_3 \tilde\phi(h_3)\right]^2 \ee

The first factor takes into account that the singlet site is white. The 
second factor gives the probability that both neighbors of the down site 
flip up under a $p_1$-process.

\subsection{$P_{\uparrow\downarrow\uparrow K}$}

$P_{\uparrow\downarrow\uparrow K}$ may be obtained on similar lines as 
$P_{\uparrow\downarrow\uparrow H}$. We get,

\be P_{\uparrow\downarrow\uparrow K}(h_a)= 2 \{1-p_0(h_a)\} 
\int_{-\infty}^{h_a} dh^{\prime} \tilde\phi(-h^\prime) 
\int_{-\infty}^{h^\prime} dh^{\prime\prime} 
P_{\downarrow\downarrow}(3\downarrow|2\downarrow; h^{\prime\prime}) 
\phi(-h^{\prime\prime}-2|J|) \ee

\subsection{$P_{\uparrow\downarrow\uparrow L}$}

The fraction of singlets $P_{\uparrow\downarrow\uparrow F}$ that are 
white at their creation but black at $h_a$ is given by,

\be P_{\uparrow\downarrow\uparrow L}= P_{\uparrow\downarrow\uparrow F} - 
P_{\uparrow\downarrow\uparrow J} -  P_{\uparrow\downarrow\uparrow K} \ee

\subsection{$P_{\uparrow\uparrow\uparrow M}$}

We now turn to the destruction i.e. the disappearance of a singlet due 
to the down spin flipping up under a $p_2$-process. We start with the 
destruction of singlets associated with $P_{\uparrow\downarrow\uparrow 
B}$. Taking into account the inequalities $h_2-2|J| < h_3 < h_2$ that 
require site-2 to be down and a black site when site-3 flips up, we get

\be P_{\uparrow\uparrow\uparrow M}(h_a)= 2 \int_{-\infty}^{h_a} 
dh^\prime \tilde\phi(-h^\prime+2|J|) 
\int_{h^\prime-4|J|}^{h^\prime-2|J|} dh^{\prime\prime} 
\phi(-h^{\prime\prime}-2|J|)  
P_{\downarrow\downarrow}(3\downarrow|2\downarrow; h^{\prime\prime}) \ee

\subsection{$P_{\uparrow\uparrow\uparrow N}$}

We now turn to the destruction of singlets associated with 
$P_{\uparrow\downarrow\uparrow E}$. These singlets were created under 
the condition $h_2+2|J| > h_3 > h_2$. They will be destroyed at 
$h_2-2|J|+h^\prime=0$ with the probability $\tilde\phi(-h^\prime+2|J|)$. 
Thus the cumulative fraction of the destroyed singlets at $h_a$ is 
given by,

\be P_{\uparrow\uparrow\uparrow N}(h_a)= 2 \int_{-\infty}^{h_a} 
dh^\prime \tilde\phi(-h^\prime+2|J|) 
\int_{h^\prime-4|J|}^{h^\prime-2|J|} dh^{\prime\prime} 
\tilde\phi(-h^{\prime\prime}) \ee

\subsection{$P_{\uparrow\uparrow\uparrow O}$}

We now consider the destruction of singlets associated with 
$P_{\uparrow\downarrow\uparrow C}$. Recall that 
$P_{\uparrow\downarrow\uparrow C}$ are white at creation. A fraction 
$P_{\uparrow\downarrow\uparrow I}$ of these become black, say at 
$h^\prime$ i.e. these can flip up at $h^{\prime}$ under a $p_0$-process 
if they were to have both neighbors down. However they have both 
neighbors up. Therefore they would flip up at applied field 
$h_a=h^\prime+4|J|$. Thus we get,

\be P_{\uparrow\uparrow\uparrow O}=P_{\uparrow\downarrow\uparrow 
I}(h_a-4|J|) \ee

\subsection{$P_{\uparrow\uparrow\uparrow P}$}

$P_{\uparrow\uparrow\uparrow P}$ represents the destruction of singlets 
associated with $P_{\uparrow\downarrow\uparrow F}$. This is 
similar to the preceding case because 
$P_{\uparrow\downarrow\uparrow F}$ are also white at creation. 
Following similar reasoning as used in the preceding section,

\be P_{\uparrow\uparrow\uparrow P}=P_{\uparrow\downarrow\uparrow 
L}(h_a-4|J|) \ee

\subsection{$P_{\uparrow\downarrow\uparrow Q}$}

Refer to figure (15) and the definition of $P_{\uparrow\downarrow 
\uparrow Q}$. We have $h_1<h_2$, $h_5<h_4$. The screening property of 
doublets ensures that the evolution of sites 2, 3, and 4 is uninfluenced 
by sites 1 and 5. Also, $h_3>h_4>h_2$ for site-3 to have flipped up 
before sites 2 and 4 ($h_3>h_2>h_4$ would make an equal contribution). 
Now site-2 flips up at $h_a$ and site-3 flips down. Therefore, 
$h_2+h_a=0$, and $h_3-2|J|+ha<0$. Consequently $h_3-h_2<2|J|$, 
$h_4<h_3<h_2+2|J|$, and $h_2<h_4<h_2+2|J|$. The probability that site-3 
would flip down when site-2 flips up is given by

\be P_{\uparrow\downarrow\uparrow Q}(h_a)=2 
\int_{-\infty}^{h_a}dh^\prime 
\phi(-h^\prime) P_{\downarrow\downarrow}(1\downarrow|2\downarrow; 
h^\prime) \int_{h^\prime-2|J|}^{h^\prime} dh^{\prime\prime} 
\phi(-h^{\prime\prime}) 
P_{\downarrow\downarrow}(5\downarrow|4\downarrow; h^{\prime\prime}) 
\int_{-h^{\prime\prime}}^{-h^\prime+2|J|} 
dh^{\prime\prime\prime}\phi(-h^{\prime\prime\prime}-2|J|) \ee

The limits on the integrals were discussed just before the equation. 
Note that one can use the quenched field at a site as a variable of 
integration or equivalently the applied field at which the site in 
question flips up. We have written the integrals in terms of the applied 
fields $h^{\prime\prime\prime}, h^{\prime\prime}, h^{\prime}$ at which 
sites 3 , 4, and 2 flip up respectively ( $h^{\prime\prime\prime} 
<h^{\prime\prime} <h^{\prime}$). The explanation of the integrands is as 
follows. The integrand in the last integral is the probability that 
site-3 flipped up by a $p_0$-process at $h_3+2|J|+ 
h^{\prime\prime\prime}=0$. The second integrand is the probability that 
site-4 flips up at $h_4+h^{\prime\prime}=0$ and site-5 is down. 
Similarly the first integrand is the probability that site-2 flips up at 
$h^\prime$ and site-1 is down.

\subsection{$P_{\uparrow\downarrow\uparrow R}$}

Refer to figure (16) and the definition of $P_{\uparrow\downarrow 
\uparrow R}$. Our object is to calculate the probability that site-3 
flips down when site-2 and site-4 are up. This happens only if site-3 
flips up after site-1 and site-5, i.e. sites 2, 3, 4 form a string of 
three down spins bordered by up spins at 1 and 5 just before 3 flips up. 
Because sites 2 and 4 are adjacent to up spins the distribution of $h_2$ 
and $h_4$ is the {\em{a posteriori}} $\tilde\phi(h_2)$ and 
$\tilde\phi(h_4)$ respectively. Also $h_3>h_2-2|J|$ and $h_3>h_4-2|J|$ 
because 3 flips up before 2 and 4. We assume $h_4>h_2$ and multiply our 
result by a factor of 2 to include the case $h_4<h_2$. Thus 
$h_4-2|J|<h_3<h_2$, $h_4-h_2<2|J|$, and $h_2<h_4<h_2+2|J|$. The 
probability that site-3 flips down when site-2 flips up is given by

\be P_{\uparrow\downarrow\uparrow R}(h_a)=2 
\int_{-\infty}^{h_a}dh^\prime 
\tilde\phi(-h^\prime+2|J|) 
\int_{h^\prime-2|J|}^{h^\prime} dh^{\prime\prime} 
\tilde\phi(-h^{\prime\prime} +2|J|) 
\int_{-h^{\prime\prime}}^{-h^\prime+2|J|} 
dh^{\prime\prime\prime}\phi(-h^{\prime\prime\prime}-2|J|) \ee

The integrand of the first and the second integral is the probability 
that site-2 and site-4 flip up by a $p_2$-process at $h^\prime$ and 
$h^{\prime\prime}$ respectively. The last integrand is the probability 
that site-3 flips up by a $p_0$- process at $h^{\prime\prime\prime}$.

\subsection{$P_{\uparrow\downarrow\uparrow S}$}

In figure(17) site-2 or site-4 could flip up first. Let us consider the 
case when site-2 flips up first i.e. $h_2>h_4-2|J|$. 
$P_{\uparrow\downarrow \uparrow S}$ is zero unless $h_2<h_3<h_4$ and 
$h_4-2|J|<h_2<h_4$. Therefore the probability that a new singlet is 
created at site-3 when site-4 flips up is given by

\be P_{\uparrow\downarrow\uparrow S}(h_a)=2 
\int_{-\infty}^{h_a}dh^\prime 
\tilde\phi(-h^\prime+2|J|) 
\int_{h^\prime-2|J|}^{h^\prime} dh^{\prime\prime} 
\phi(-h^{\prime\prime}) 
P_{\downarrow\downarrow}(1\downarrow|2\downarrow; h^{\prime\prime}) 
\int_{-h^{\prime\prime}}^{-h^\prime+2|J|} 
dh^{\prime\prime\prime}\phi(-h^{\prime\prime\prime}-2|J|) \ee

The last integrand has the same interpretation as in the previous object 
$P_{\uparrow\downarrow\uparrow R}$. The second integrand is the 
probability that site-1 is down when site-2 flips up by a $p_1$-process 
at $h^{\prime\prime}$. Finally the first integrand is the probability 
that site-4 flips up by a $p_2$-process at $h^\prime$

\subsection{$P_{\uparrow\downarrow\uparrow T}$}

Site-4 would flip up before site-2 in figure(17) if $h_4>h_2+2|J|$. 
Clearly $h_3>h_2$. Site-3 would flip down when site-2 flips up if 
$h_3<h_2+2|J|$. Thus $h_3$ and $h_4$ lie in the range 
$h_4-2|J|<h_3<h_2+2|J|$ and $h_2+2|J|<h_4<h_2+4|J|$. The probability 
that site-3 flips down when site-2 flips up is given by

\be P_{\uparrow\downarrow\uparrow T}(h_a)=2 \int_{-\infty}^{h_a} 
dh^{\prime} \phi(-h^\prime) 
P_{\downarrow\downarrow}(1\downarrow|2\downarrow; h^{\prime}) 
\int_{h^\prime-2|J|}^{h^\prime}dh^{\prime\prime} 
\tilde\phi(-h^{\prime\prime}+2|J|) 
\int_{-h^{\prime\prime}}^{-h^\prime+2|J|} 
dh^{\prime\prime\prime}\phi(-h^{\prime\prime\prime}-2|J|) \ee

The integrand is similar to that in $P_{\uparrow\downarrow\uparrow S}$ 
except that the first integral is for $h_2$ and the second for $h_4$.

\subsection{$P_{\uparrow\uparrow\uparrow U}$}

$P_{\uparrow\uparrow\uparrow U}$ is associated with the destruction of 
objects characterized by $P_{\uparrow\downarrow\uparrow Q}$. The 
destruction can be analyzed on similar lines as their creation except 
that we now have the inequalities $h_2<h_4<h_3$ and $h_3-2|J|<h_2<h_3$. 
The probability that site-3 in $P_{\uparrow\downarrow\uparrow Q}$ flips 
for the third time is given by

\be P_{\uparrow\uparrow\uparrow U}(h_a)=2 
\int_{-\infty}^{h_a} dh^{\prime} \phi(-h^\prime+2|J|) 
\int_{h^\prime-2|J|}^{h^\prime} dh^{\prime\prime} 
\phi(-h^{\prime\prime})  
P_{\downarrow\downarrow}(1\downarrow|2\downarrow; h^{\prime\prime})
\int_{h^\prime-2|J|}^{h^{\prime\prime}} dh^{\prime\prime\prime} 
\phi(-h^{\prime\prime\prime})  
P_{\downarrow\downarrow}(5\downarrow|4\downarrow; 
h^{\prime\prime\prime}) \ee

The second and the third integrands give the probability that sites 2 
and 4 flip up at $h^{\prime\prime}$ and $h^{\prime\prime\prime}$ 
respectively. The first integrand is the probability that site-3 flips 
up by a $p_2$-process at $h^\prime$. Note that {\em{ a priori}} 
distributions $\phi(h_2),\phi(h_3),\phi(h_4)$ are used here because 
these sites have remained screened from site-1 and site-5.

\subsection{$P_{\uparrow\uparrow\uparrow V}$}

$P_{\uparrow\uparrow\uparrow V}$ is associated with the destruction of 
$P_{\uparrow\downarrow\uparrow R}$. The inequalities that govern 
$P_{\uparrow\uparrow\uparrow V}$ are $h_2<h_4<h_3+2|J|$ and 
$h_3<h_2<h_3+2|J|$. Thus the probability that the down spin in 
$P_{\uparrow\downarrow\uparrow R}$ flips for the third time is

\be P_{\uparrow\uparrow\uparrow V}(h_a)=2 
\int_{-\infty}^{h_a} dh^{\prime} \phi(-h^\prime+2|J|) 
\int_{h^\prime-2|J|}^{h^\prime} dh^{\prime\prime} 
\tilde\phi(-h^{\prime\prime}+2|J|)
\int_{h^\prime-2|J|}^{h^{\prime\prime}} dh^{\prime\prime\prime}  
\tilde\phi(-h^{\prime\prime\prime}+2|J|) \ee

The second and the third integrands give the probability that site-2 and 
site-4 flip up by a $p_2$-process at $h^{\prime\prime}$ and 
$h^{\prime\prime\prime}$ respectively. The first integrand is the 
probability that site-3 flips up by a $p_2$-process at $h^\prime$.

\subsection{$P_{\uparrow\uparrow\uparrow W}$}

$P_{\uparrow\uparrow\uparrow W}$ is associated with the destruction of 
$P_{\uparrow\downarrow\uparrow S}$. The appropriate inequalities for  
$P_{\uparrow\uparrow\uparrow W}$ are $h_3<h_4<h_3+2|J|$ 
and $h_4-2|J|<h_2<h_3$. The probability that the down spin in question  
flips for the third time is

\be P_{\uparrow\uparrow\uparrow W}(h_a)=2 
\int_{-\infty}^{h_a} dh^{\prime} \phi(-h^\prime+2|J|) 
\int_{h^\prime-2|J|}^{h^\prime} dh^{\prime\prime} 
\tilde\phi(-h^{\prime\prime}+2|J|)
\int_{h^\prime-2|J|}^{h^{\prime\prime}} dh^{\prime\prime\prime}  
\phi(-h^{\prime\prime\prime}) 
P_{\downarrow\downarrow}(1\downarrow|2\downarrow; 
h^{\prime\prime\prime})
\ee

The first integrand is the probability that site-3 flips up by a 
$p_2$-process at $h^\prime$. The second and third integrands account for 
site-4 and site-2 respectively.

\subsection{$P_{\uparrow\uparrow\uparrow X}$}

$P_{\uparrow\uparrow\uparrow X}$ gives the fraction of singlets 
associated with {$P_{\uparrow\downarrow\uparrow T}$ that are destroyed 
at $h_a$. These may be calculated in a similar manner as in the 
preceding case. We now have the inequalities $h_2+2|J|<h_4<h_3+2|J|$ and 
$h_3-2|J|<h_2<h_3$. Therefore the probability that the down spin in 
{$P_{\uparrow\downarrow\uparrow T}$ flips up is given by,

\be P_{\uparrow\uparrow\uparrow X}(h_a)=2 
\int_{-\infty}^{h_a} dh^{\prime} \phi(-h^\prime+2|J|) 
\int_{h^\prime-2|J|}^{h^\prime} dh^{\prime\prime} 
\phi(-h^{\prime\prime})
P_{\downarrow\downarrow}(1\downarrow|2\downarrow; 
h^{\prime\prime})
\int_{h^\prime-2|J|}^{h^{\prime\prime}} dh^{\prime\prime\prime}  
\tilde\phi(-h^{\prime\prime\prime}+2|J|) 
\ee

The last two integrands pertain to sites 2 and 4 respectively, and the 
first to site-3.

\section{Magnetization on lower hysteresis loop}

We are now in a position to write the magnetization on the lower half of 
the hysteresis loop. 

\be m(h_a)=1-2P_{\downarrow}(h_a) \ee

where $P_{\downarrow}(h_a)$ is the probability that a randomly chosen 
site on the chain is down at applied field $h_a$. A randomly chosen site 
on the chain can be characterized by the number $n$ of up neighbors it 
has ($n=0,1,2$). Thus we can write,

\be P_{\downarrow}(h_a) =P_{\downarrow\downarrow\downarrow}(h_a) + 
P_{\downarrow\downarrow\uparrow}(h_a) + 
P_{\uparrow\downarrow\downarrow}(h_a) + 
P_{\uparrow\downarrow\uparrow}(h_a) \ee

The first term corresponds to $n=0$, the next two terms that are equal 
by symmetry correspond to $n=1$, and the last term corresponds to $n=2$. 
We obtained the first three terms on the right-hand-side with relative 
ease in section (V). Surprisingly the evaluation of the last term i.e. 
$P_{\uparrow\downarrow\uparrow}(h_a)$ proved rather tedious requiring 
the calculation of 24 terms as a pre-requisite 
[$P_{\uparrow\downarrow\uparrow A}(h_a)$ to $P_{\uparrow\uparrow\uparrow 
X}(h_a)$]. We have (so far) not found a simpler method to calculate 
$P_{\uparrow\downarrow\uparrow}(h_a)$ in spite of much effort and 
thought. Putting all terms together we get,

\bea P_{\downarrow}(h_a) = 
 P_{\downarrow\downarrow\downarrow}(h_a)+ 
 P_{\downarrow\downarrow\uparrow}(h_a)+ 
 P_{\uparrow\downarrow\downarrow}(h_a)+ 
 P_{\uparrow\downarrow\uparrow A}(h_a)+ 
 P_{\uparrow\downarrow\uparrow D}(h_a) \nonumber \\ 
  - P_{\uparrow\uparrow\uparrow M}(h_a)
  - P_{\uparrow\uparrow\uparrow N}(h_a) 
  - P_{\uparrow\uparrow\uparrow 0}(h_a)
  - P_{\uparrow\uparrow\uparrow P}(h_a) \nonumber \\
  + P_{\uparrow\downarrow\uparrow Q}(h_a) 
  + P_{\uparrow\downarrow\uparrow R}(h_a) 
  + P_{\uparrow\downarrow\uparrow S}(h_a) 
  + P_{\uparrow\downarrow\uparrow T}(h_a) \nonumber \\
  - P_{\uparrow\uparrow\uparrow U}(h_a)- 
 P_{\uparrow\uparrow\uparrow V}(h_a)- 
 P_{\uparrow\uparrow\uparrow W}(h_a)- 
 P_{\uparrow\uparrow\uparrow X}(h_a) \eea

We note that several objects that we calculated in the preceding section 
do not appear explicitly in the above equation e.g. $ 
P_{\uparrow\downarrow\uparrow B}(h_a)$ does not appear explicitly in 
equation (89). However it is necessary to calculate $ 
P_{\uparrow\downarrow\uparrow B}(h_a)$ because it is needed in the 
calculation of $ P_{\uparrow\downarrow\uparrow M}(h_a)$ and $ 
P_{\uparrow\downarrow\uparrow O}(h_a)$ that appear in the final formula. 
Similar remarks apply to other terms that were calculated but do not 
appear explicitly in equation (89).

The magnetization in increasing field is obtained by substituting 
equation (89) in equation (87). The magnetization in decreasing field on 
the upper half of the hysteresis loop is given by symmetry,

\be m^u(h_a)=-m(-h_a) \ee

In the next section we compare the theoretical result against numerical 
simulations of the model in selected cases. We consider a uniform 
bounded distribution of the quenched field as well as a Gaussian 
distribution. As may be anticipated, the agreement between theory and 
numerical simulation is quite good.

\section{Comparison with simulations and concluding remarks}

Simulations of the model have played an important role in the analysis 
presented here. Although an exact analytic result has to be necessarily 
in agreement with the simulations within numerical errors but arguments 
based on conditional probabilities can be subtle and prone to errors. 
Therefore at each step of the analysis, we devised a simulation of the 
model to yield the probability of the event being calculated. 
Occasionally the two would not match in the first instance necessitating 
a rethink of the analysis and locating the error in the argument. Thus 
each of the theoretical expression in the preceding sections was 
verified by simulation of the model for a bounded distribution of the 
random field with half-width $\Delta=1.25 |J|$ and a Gaussian 
distribution with standard deviation $\sigma=.5 |J|$. The comparison 
between the theoretical hysteresis loops and those obtained by 
simulation was shown already in section III. Here we show the comparison 
for a few other quantities that enter the expression for the hysteresis 
loops.

We begin with the probability per site of a doublet on the lower half of 
the hysteresis loop. Figure (18) shows the theoretical expression for 
$P_{\downarrow\downarrow}(h_a)$ for the Gaussian distribution with the 
corresponding data from numerical simulation superimposed on it. The fit 
is so close that the two are indistinguishable on the scale of the 
figure. Figure (19) shows similar comparison for a uniform distribution 
with $\Delta=1.25$.  In each of the following figures data from the 
corresponding numerical simulation has been plotted along side the 
theoretical expression. Even numbered figures are for the Gaussian 
distribution and the odd numbered figures for the uniform distribution.  
In some cases the match between theory and simulation is so good that 
there appears to be only a single curve in the figure. In other cases ( 
when the probability of the event is relatively small and finite size 
corrections are larger) we can barely make out that there are two curves 
that almost lie on each other. Figure(20) and figure (21) show the 
result for $P_{\downarrow\downarrow\downarrow}(h_a)$ for the Gaussian 
and the uniform distribution respectively. Results for 
$P_{\uparrow\downarrow\downarrow}(h_a) + 
P_{\downarrow\downarrow\uparrow}(h_a)$ are shown in figures (22) and 
(23) for Gaussian and uniform distributions respectively. Figure (24) 
shows the comparison between theory and simulation for 
$P_{\uparrow\downarrow\uparrow A}(h_a)$ in the case of Gaussian 
distribution. Figure (25) shows similar comparison for 
$P_{\uparrow\downarrow\uparrow D}(h_a)$ for the uniform distribution. 
Figure (26) is for $P_{\uparrow\uparrow\uparrow M}(h_a)$ for the 
Gaussian distribution. Figure (27) is for $P_{\uparrow\uparrow\uparrow 
N}(h_a)$ for the rectangular distribution. Figures (28) and (29) each 
contain two objects. Figure (28) shows $P_{\uparrow\downarrow\uparrow 
Q}(h_a)$ and $P_{\uparrow\uparrow\uparrow U}(h_a)$ for Gaussian 
distribution. Figure (29) shows $P_{\uparrow\downarrow\uparrow R}(h_a)$ 
and $P_{\uparrow\uparrow\uparrow V}(h_a)$ for uniform distribution.

In conclusion we have obtained the zero-temperature hysteresis loop of a 
one dimensional anti-ferromagnetic random field Ising model in the case 
when the driving field varies from $-\infty$ to $\infty$ and back to 
$-\infty$ infinitely slowly. The problem is simple to state but 
difficult to solve. The theoretical result for the hysteresis loop 
involves integrals that have to be evaluated numerically in most cases. 
We have shown that our results fit numerical simulations of the model 
quite well. However, several aspects of the problem still remain 
unsolved. For example we have obtained the hysteresis loop when the 
driving field takes the system from one saturated state ($h_a=-\infty$) 
to another ($h_a=\infty$). In this case we have a complete knowledge of 
the statistical history of the system i.e. if a site is up at $h_a$, we 
know the relative probability of different sequence of events that 
result in this site being up. We are not in a position (so far) to 
obtain the hysteretic response of the system starting from an arbitrary 
initial state. It would be interesting to have an analytic solution of 
the problem in higher dimensions as well. Numerical simulations suggest 
that the anti-ferromagnetic hysteresis loops in higher dimensions have 
several plateaus for low values of $\Delta$ and $\sigma$ as compared 
with $|J|$. Exact analytic solutions of problems with quenched disorder 
are uncommon in statistical mechanics.  One may even ask if they are 
worth the effort that has to be put in trying to obtain them. However 
exact solutions are intellectually satisfying and provide a framework 
for understanding a wide class of complex phenomena. We hope there will 
be more progress in this direction in the future.


\begin{figure}[p] 
\includegraphics[width=.75\textwidth,angle=0]{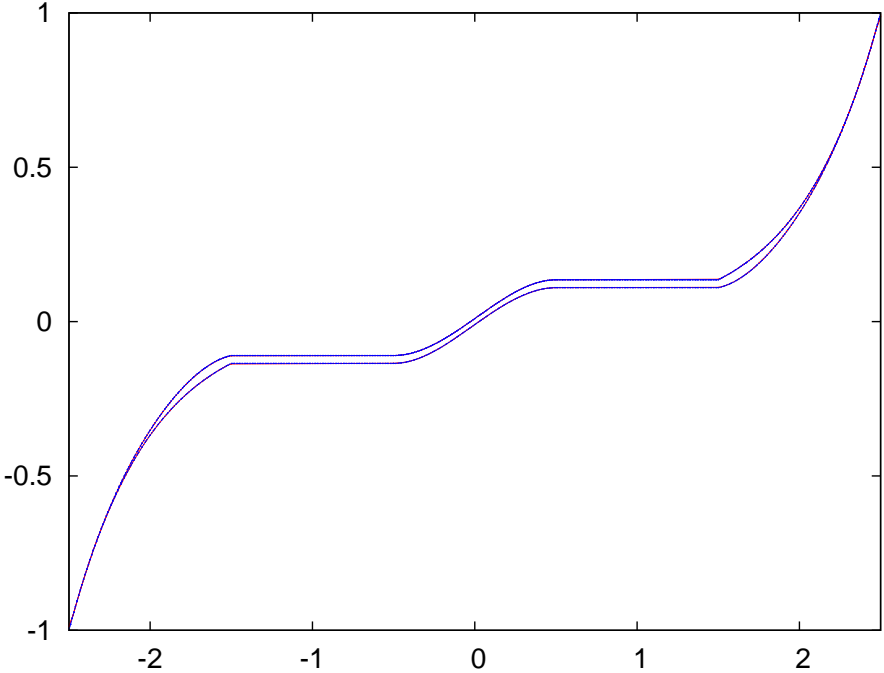} 
\caption{Hysteresis loop for an anti-ferromagnetic random-field Ising 
model with $J=-1$ and $\Delta=0.5$ (see text). The $x$-axis shows the 
applied field and the $y$-axis magnetization per spin. As $|J| \le 
\Delta$, each half of the hysteresis loop comprises three ramps 
separated by two plateaus. The lower half of the loop shows 
magnetization in increasing field and the upper half in decreasing 
field. A theoretical expression has been superimposed on the numerical 
data.} \label{fig1} \end{figure}

\begin{figure}[p] 
\includegraphics[width=.75\textwidth,angle=0]{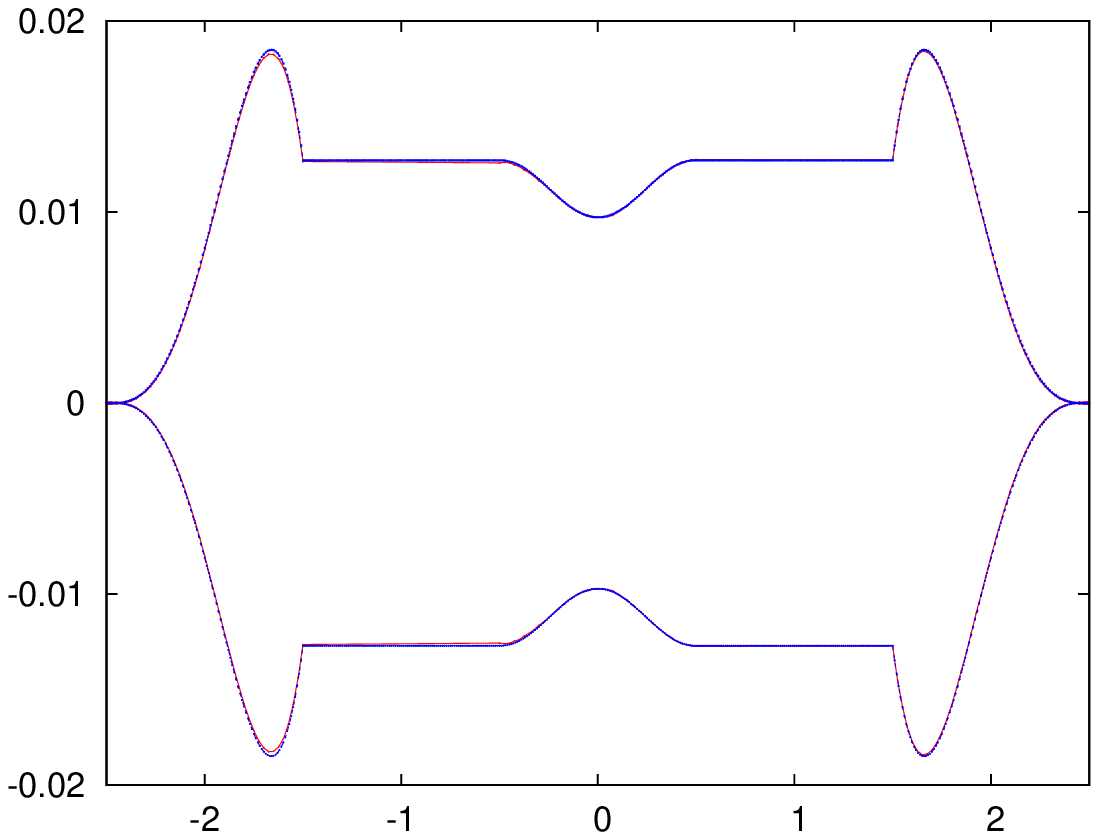}
\caption{ A magnified view of the theoretical and simulation hysteresis 
loops for $\Delta=0.5$ where the $y$-axis shows the magnetization in 
increasing and decreasing field as measured from the average of the 
magnetization on the lower and the upper half of the hysteresis loop in 
figure ~\ref{fig1} at corresponding applied field.}
\label{fig2} \end{figure}

\begin{figure}[p] 
\includegraphics[width=.75\textwidth,angle=0]{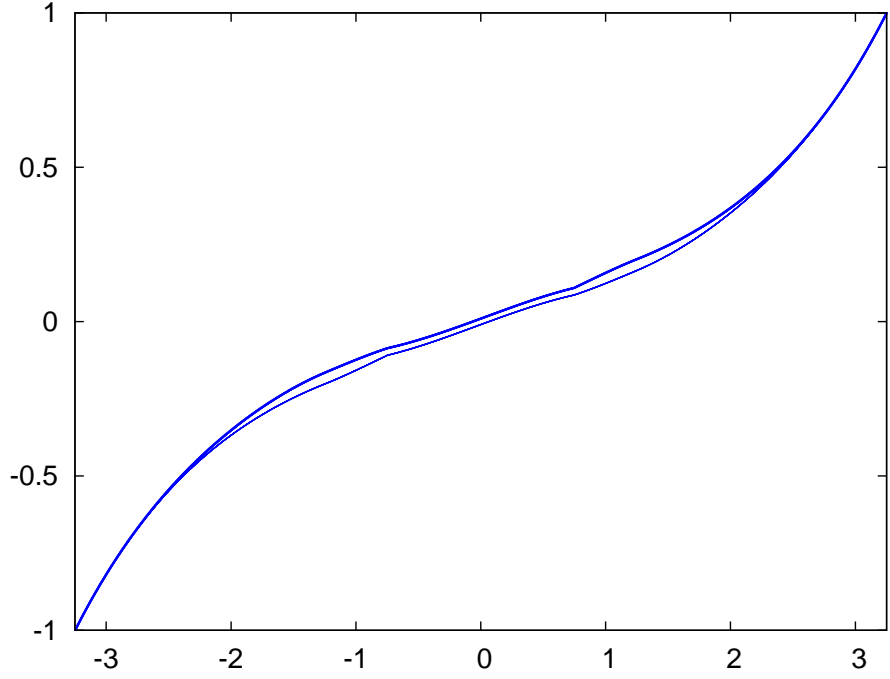}
\caption{Hysteresis loop for an anti-ferromagnetic random-field Ising 
model with $J=-1$ and $\Delta=1.25$ (see text). The $x$-axis shows the 
applied field and the $y$-axis magnetization per spin. As $|J| > 
\Delta$, the plateaus of figure ~\ref{fig1} disappear and the three 
ramps merge into each other. A theoretical expression has been 
superimposed on the numerical data.}
\label{fig3} \end{figure}

\begin{figure}[p] 
\includegraphics[width=.75\textwidth,angle=0]{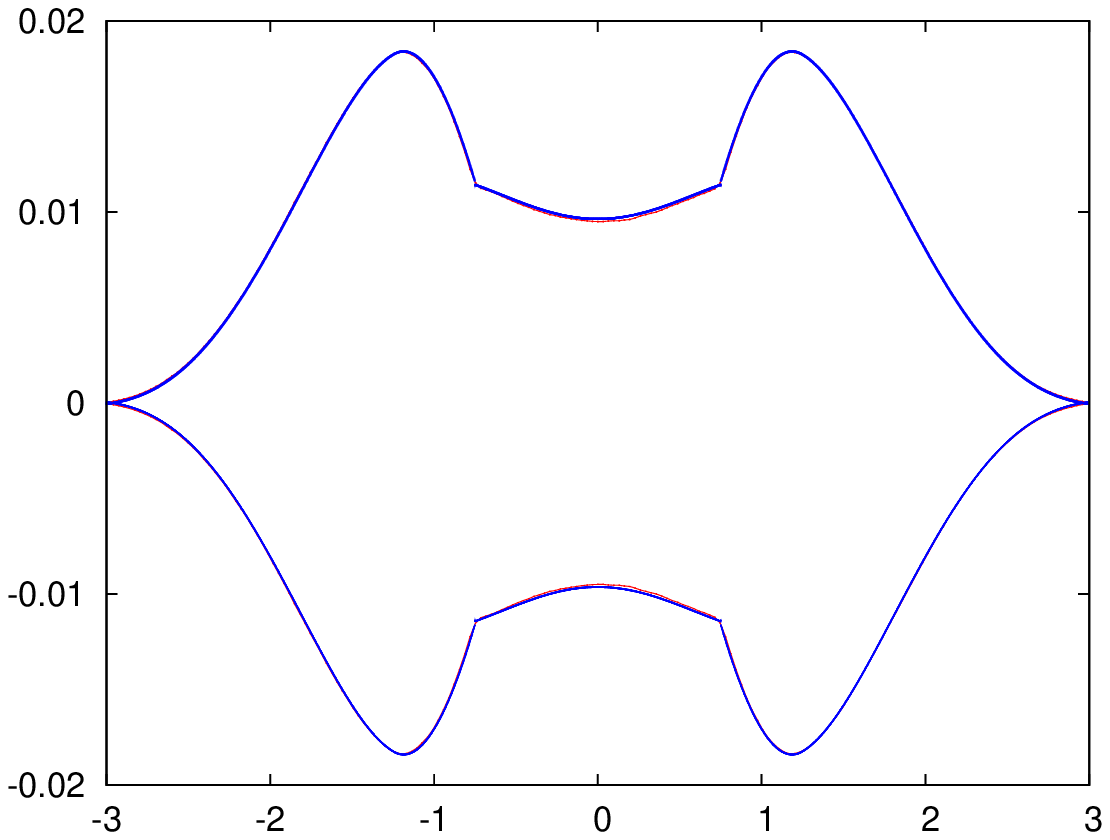}
\caption{ Theoretical and simulation hysteresis loops for $\Delta=1.25$ 
where the magnetization along increasing and decreasing field is 
measured from the average of the magnetization on the lower and the 
upper half of the hysteresis loop in figure ~\ref{fig3} at corresponding 
applied field.}
\label{fig4} \end{figure}

\begin{figure}[p] 
\includegraphics[width=.75\textwidth,angle=0]{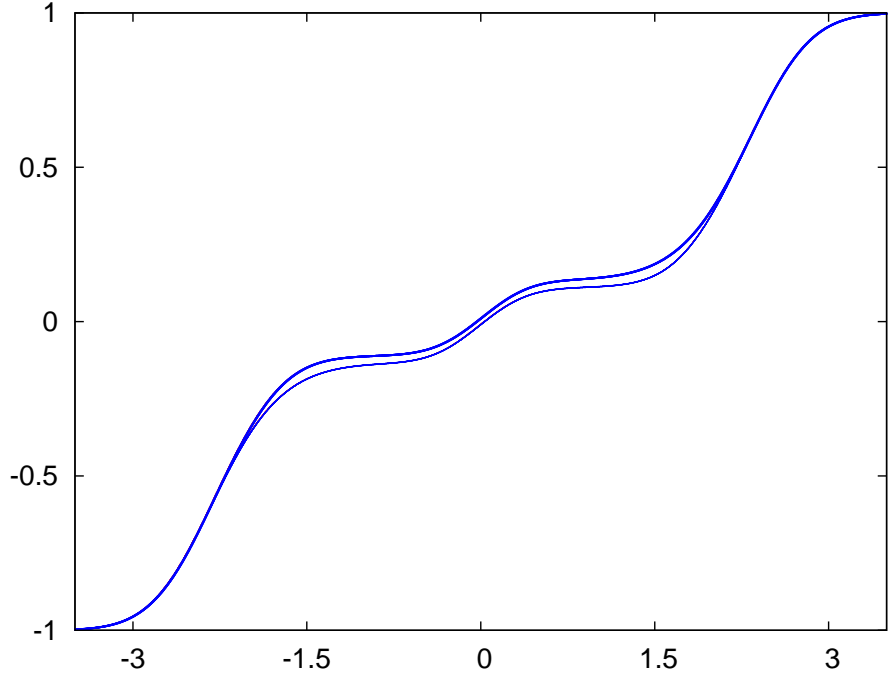}
\caption{Hysteresis loop for an anti-ferromagnetic random-field Ising 
model with $J=-1$ and $\sigma=0.5$ (see text). Notice the approximate 
similarity with the hysteresis loop in figure ~\ref{fig1} but the 
absence of sharp ramps and plateaus. For a Gaussian distribution the 
three ramps merge into each other for any value of $\sigma$ although 
this is less pronounced at smaller values of $\sigma$. A theoretical 
expression has been superimposed on the numerical data.} \label{fig5} 
\end{figure}

\begin{figure}[p] 
\includegraphics[width=.75\textwidth,angle=0]{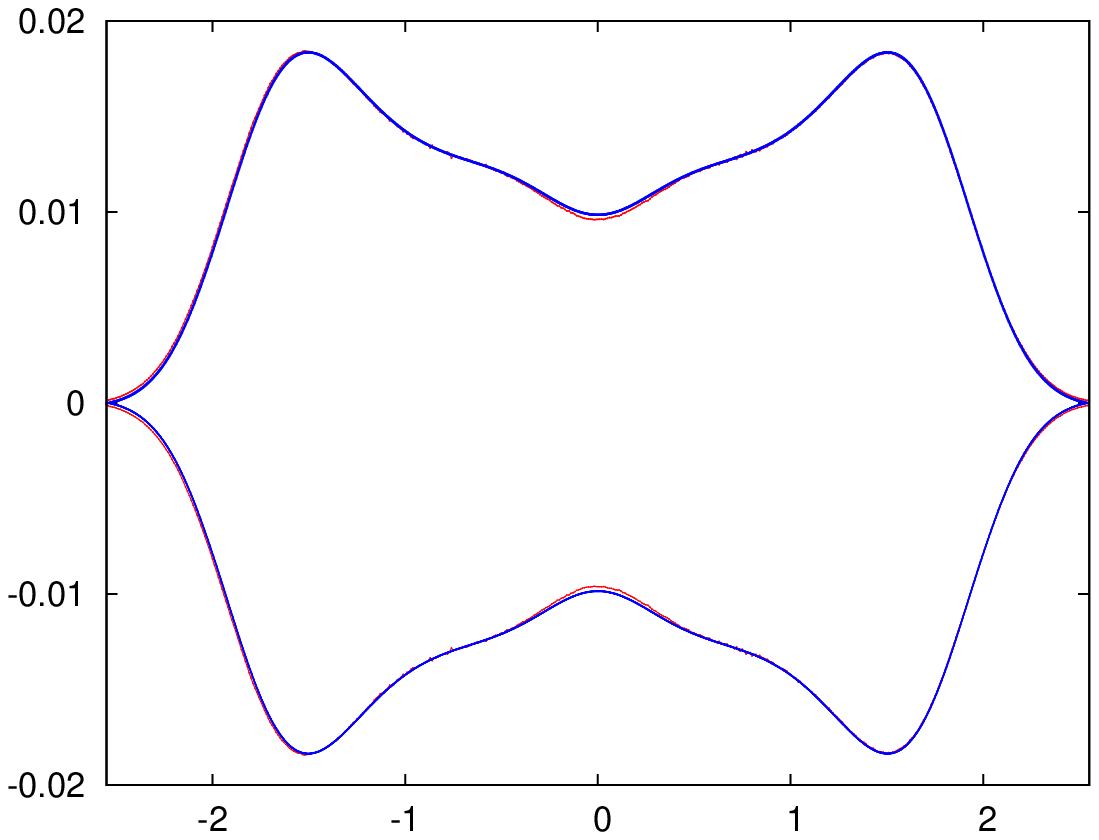}
\caption{ Magnified theoretical and simulation hysteresis loops for 
$\sigma=0.5$ where the magnetization along increasing and decreasing 
field is measured from the average of the magnetization on the lower and 
the upper half of the hysteresis loop in figure ~\ref{fig5} at 
corresponding applied field.}
\label{fig6} \end{figure}

\thispagestyle{empty}
\setlength{\unitlength}{1cm}
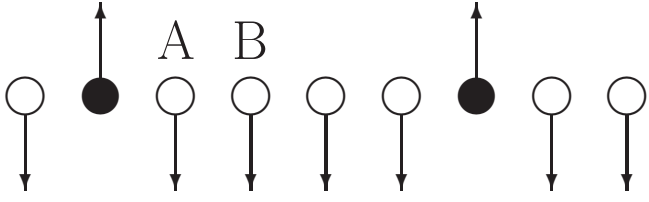
\begin{figure}[p]
\begin{picture}(15.,2.)
\thicklines
\put(0,1){\circle{.5}} \put(0,.75){\vector(0,-1){1}}
\put(1,1){\circle*{.5}} \put(1,1.25){\vector(0,1){1}}
\put(2,1){\circle{.5}} \put(2,.75){\vector(0,-1){1}}
\put(3,1){\circle{.5}} \put(3,.75){\vector(0,-1){1}}
\put(4,1){\circle{.5}} \put(4,.75){\vector(0,-1){1}}
\put(5,1){\circle{.5}} \put(5,.75){\vector(0,-1){1}}
\put(6,1){\circle*{.5}} \put(6,1.25){\vector(0,1){1}}
\put(7,1){\circle{.5}} \put(7,.75){\vector(0,-1){1}}
\put(8,1){\circle{.5}} \put(8,.75){\vector(0,-1){1}}
\put(1.75,1.5){\huge{A}}
\put(2.75,1.5){\huge{B}}
\end{picture}
\vspace{1.5cm}
\caption{Spins on Ramp-I in an applied field $-2|J|-h$.
Filled circles show sites with quenched field $h_{i} > h$.
The probability per site of a doublet (two adjacent down spins)
such as AB is equal to $e^{-2p}$, where p is the fraction of
filled circles on the infinite lattice.}
\label{fig7}
\end{figure}

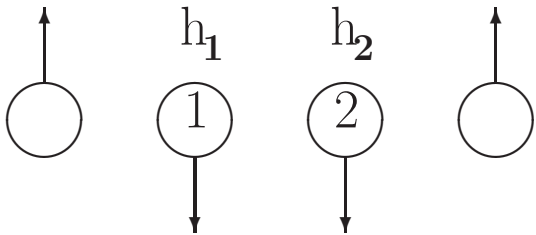
\begin{figure}[p]
\begin{picture}(15.,6.)
\thicklines
\put(1,-2){\circle{1}}
\put(.85,-2.1){\huge{}} \put(1,-1.5){\vector(0,1){1}}
\put(3,-2){\circle{1}}
\put(2.85,-2.1){\huge{1}} \put(3,-2.5){\vector(0,-1){1}}
\put(5,-2){\circle{1}}
\put(4.85,-2.1){\huge{2}} \put(5,-2.5){\vector(0,-1){1}}
\put(7,-2){\circle{1}}
\put(6.85,-2.1){\huge{}} \put(7,-1.5){\vector(0,1){1}}
\put(2.8,-1){\huge{h}}
\put(3.1,-1.2){\bf{\Large{1}}}
\put(4.8,-1){\huge{h}}
\put(5.1,-1.2){\bf{\Large{2}}}
\end{picture}
\vspace{6cm}
\caption{A doublet on Plateau-I: $h_{1}$ and $h_{2}$ are the quenched 
random fields on the doublet sites 1 and 2 respectively.}
\label{fig8}
\end{figure}
\begin{figure}[p]
\begin{picture}(15,4)
\thicklines
\put(1,1){\circle{1}}
\put(3,1){\circle{1}}
\put(5,1){\circle{1}}
\put(7,1){\circle{1}}
\put(9,1){\circle{1}}
\put(11,1){\circle{1}}
\put(13,1){\circle{1}}
\put(1,1.5){\vector(0,1){1}}
\put(3,.5){\vector(0,-1){1}}
\put(5,.5){\vector(0,-1){1}}
\put(7,1.5){\vector(0,1){1}}
\put(9,.5){\vector(0,-1){1}}
\put(11,.5){\vector(0,-1){1}}
\put(13,1.5){\vector(0,1){1}}
\put(-.3,-1){\dashbox{0.2}(4,4)}
\put(4.3,-1){\dashbox{0.2}(5.5,4)}
\put(10.3,-1){\dashbox{0.2}(4,4)}
\put(2.8,.8){\huge{1}}
\put(4.8,.8){\huge{2}}
\put(6.8,.8){\huge{3}}
\put(8.8,.8){\huge{4}}
\put(10.8,.8){\huge{5}}
\put(2.8,2){\huge{h}}
\put(3.3,1.8){\bf{\Large{1}}}
\put(4.8,2){\huge{h}}
\put(5.3,1.8){\bf{\Large{2}}}
\put(6.8,-.5){\huge{h}}
\put(7.3,-.7){\bf{\Large{3}}}
\put(8.8,2){\huge{h}}
\put(9.3,1.8){\bf{\Large{4}}}
\put(10.8,2){\huge{h}}
\put(11.3,1.8){\bf{\Large{5}}}
\end{picture} 
\vspace{1.5cm}
\caption{Two adjacent doublets on Plateau-I: Each doublet
separates the lattice into two parts whose evolution histories
on Ramp-I are independent of each other. Evolutions inside
each dashed box is shielded from outside. The probability that
spin at site 3 flips up on Ramp-I is therefore equal to
$\frac{1}{3}$. Given this, the probability that the spins at
sites 1 and 5 remain down all along Ramp-I is equal to
$\frac{1}{e}$ each. The shielding property of the boxes can also
be used to determine a posteriori distribution of random fields
$h_{1}, h_{2}, h_{3}, h_{4}, \mbox{  and  } h_{5}$.}
\label{fig9}
\end{figure}
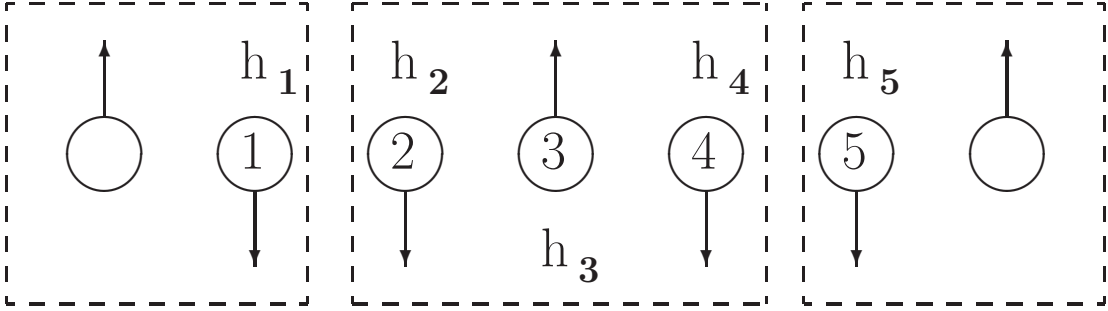

\setlength{\unitlength}{1cm}
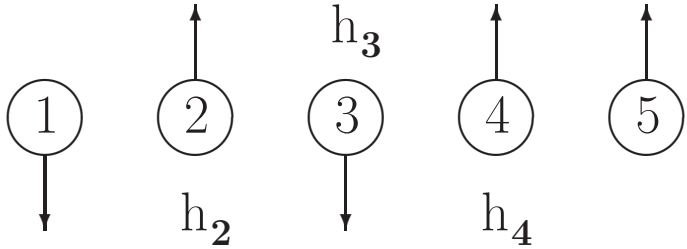
\begin{figure}[p]
\begin{picture}(9,0)
\thicklines
\put(0,-2){\circle{1}}
\put(-.15,-2.2){\huge{1}} \put(0,-2.5){\vector(0,-1){1}}
\put(2,-2){\circle{1}}
\put(1.85,-2.2){\huge{2}} \put(2,-1.5){\vector(0,1){1}}
\put(4,-2){\circle{1}}
\put(3.85,-2.2){\huge{3}} \put(4,-2.5){\vector(0,-1){1}}
\put(6,-2){\circle{1}}
\put(5.85,-2.2){\huge{4}} \put(6,-1.5){\vector(0,1){1}}
\put(8,-2){\circle{1}}
\put(7.85,-2.2){\huge{5}} \put(8,-1.5){\vector(0,1){1}}
\put(1.8,-3.5){\huge{h}}
\put(2.2,-3.7){\bf{\Large{2}}}
\put(3.8,-1){\huge{h}}
\put(4.2,-1.2){\bf{\Large{3}}}
\put(5.8,-3.5){\huge{h}}
\put(6.2,-3.7){\bf{\Large{4}}}
\end{picture}
\vspace{5cm}
\caption{A singlet (site 3) with one next nearest neighbor down 
(site 1), and one next nearest neighbor up (site 5). When the singlet 
turns up at an applied field $h_{a}$, the spin at site 2 stays up if 
$\Delta \le |J|$, but the spin at site 4 flips down if $h_{4} \le 
h_{3}$.}
\label{fig10}
\end{figure}

\setlength{\unitlength}{1cm}
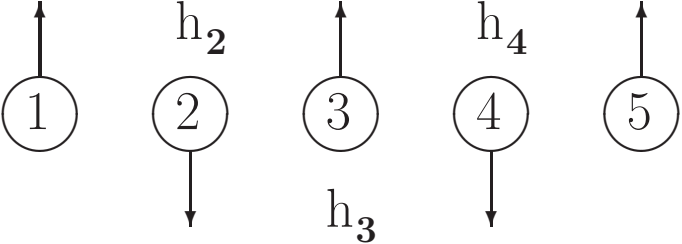
\begin{figure}[p]
\begin{picture}(9,-4)
\thicklines
\put(0,-12){\circle{1}}
\put(2,-12){\circle{1}}
\put(4,-12){\circle{1}}
\put(6,-12){\circle{1}}
\put(8,-12){\circle{1}}
\put(0,-11.5){\vector(0,1){1}}
\put(2,-12.5){\vector(0,-1){1}}
\put(4,-11.5){\vector(0,1){1}}
\put(6,-12.5){\vector(0,-1){1}}
\put(8,-11.5){\vector(0,1){1}}
\put(-.2,-12.2){\huge{1}}
\put(1.8,-12.2){\huge{2}}
\put(3.8,-12.2){\huge{3}}
\put(5.8,-12.2){\huge{4}}
\put(7.8,-12.2){\huge{5}}
\put(1.8,-11){\huge{h}}
\put(2.2,-11.2){\bf{\Large{2}}}
\put(3.8,-13.5){\huge{h}}
\put(4.2,-13.7){\bf{\Large{3}}}
\put(5.8,-11){\huge{h}}
\put(6.2,-11.2){\bf{\Large{4}}}
\end{picture}
\vspace{15cm}
\caption{ Two adjacent singlets on Plateau-I: If $h_{2} =
\mbox{min}(h_{2},h_{4})$, and $h_{3} \le h_{2}$, then the spin at site 3
will flip down when the spin at site 2 flips up on ramp-III. This process
creates a new singlet on ramp-III.}
\label{fig11}
\end{figure}
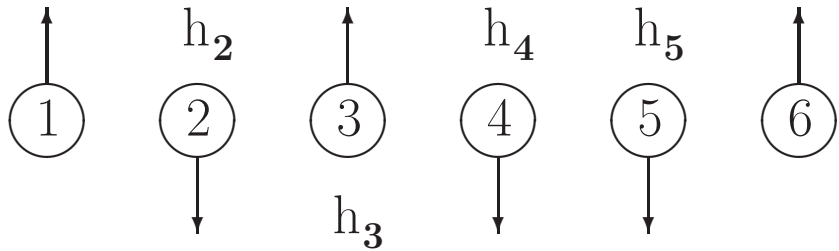
\begin{figure}
\setlength{\unitlength}{1cm}
\begin{picture}(9,0.)
\thicklines
\put(0,-2){\circle{1}}
\put(-.15,-2.2){\huge{1}} \put(0,-1.5){\vector(0,1){1}}
\put(2,-2){\circle{1}}
\put(1.85,-2.2){\huge{2}} \put(2,-2.5){\vector(0,-1){1}}
\put(4,-2){\circle{1}}
\put(3.85,-2.2){\huge{3}} \put(4,-1.5){\vector(0,1){1}}
\put(6,-2){\circle{1}}
\put(5.85,-2.2){\huge{4}} \put(6,-2.5){\vector(0,-1){1}}
\put(8,-2){\circle{1}}
\put(7.85,-2.2){\huge{5}} \put(8,-2.5){\vector(0,-1){1}}
\put(10,-2){\circle{1}}
\put(9.85,-2.2){\huge{6}} \put(10,-1.5){\vector(0,1){1}}
\put(1.8,-1){\huge{h}}
\put(2.2,-1.2){\bf{\Large{2}}}
\put(3.8,-3.5){\huge{h}}
\put(4.2,-3.7){\bf{\Large{3}}}
\put(5.8,-1){\huge{h}}
\put(6.2,-1.2){\bf{\Large{4}}}
\put(7.8,-1){\huge{h}}
\put(8.2,-1.2){\bf{\Large{5}}}
\end{picture}
\vspace{6cm}
\caption{ A singlet followed by a doublet on Plateau-I: If $h_{4} \ge
h_{5}$, and $h_{3} \le h_{2}$, then a new singlet will be created at site
3 when the spin at site 2 turns up on ramp-III.}
\label{fig12}
\end{figure}
\setlength{\unitlength}{1cm}
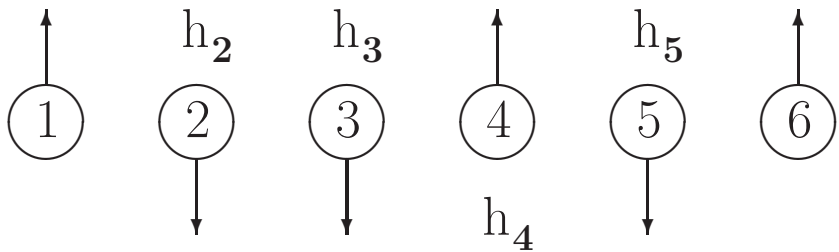
\begin{figure}
\begin{picture}(12,0)
\thicklines
\put(0,-12){\circle{1}}
\put(-.15,-12.2){\huge{1}} \put(0,-11.5){\vector(0,1){1}}
\put(2,-12){\circle{1}}
\put(1.85,-12.2){\huge{2}} \put(2,-12.5){\vector(0,-1){1}}
\put(4,-12){\circle{1}}
\put(3.85,-12.2){\huge{3}} \put(4,-12.5){\vector(0,-1){1}}
\put(6,-12){\circle{1}}
\put(5.85,-12.2){\huge{4}} \put(6,-11.5){\vector(0,1){1}}
\put(8,-12){\circle{1}}
\put(7.85,-12.2){\huge{5}} \put(8,-12.5){\vector(0,-1){1}}
\put(10,-12){\circle{1}}
\put(9.85,-12.2){\huge{6}} \put(10,-11.5){\vector(0,1){1}}
\put(1.8,-11){\huge{h}}
\put(2.2,-11.2){\bf{\Large{2}}}
\put(3.8,-11){\huge{h}}
\put(4.2,-11.2){\bf{\Large{3}}}
\put(5.8,-13.5){\huge{h}}
\put(6.2,-13.7){\bf{\Large{4}}}
\put(7.8,-11){\huge{h}}
\put(8.2,-11.2){\bf{\Large{5}}}
\end{picture}
\vspace{15cm}
\caption{ A doublet followed by a singlet on Plateau-I: If $h_{3} \ge
h_{2}$, and $h_{4} \le h_{5}$, then a new singlet will be created at site
4 when the spin at site 5 turns up on ramp-III.}
\label{fig 13}
\end{figure}
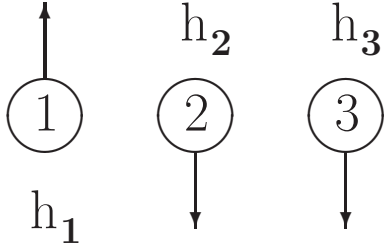
\begin{figure}
\setlength{\unitlength}{1cm}
\begin{picture}(9,0.)
\thicklines
\put(0,-2){\circle{1}}
\put(-.15,-2.2){\huge{1}} \put(0,-1.5){\vector(0,1){1}}
\put(2,-2){\circle{1}}
\put(1.85,-2.2){\huge{2}} \put(2,-2.5){\vector(0,-1){1}}
\put(4,-2){\circle{1}}
\put(3.85,-2.2){\huge{3}} \put(4,-2.5){\vector(0,-1){1}}
\put(-0.2,-3.5){\huge{h}}
\put(0.2,-3.7){\bf{\Large{1}}}
\put(1.8,-1){\huge{h}}
\put(2.2,-1.2){\bf{\Large{2}}}
\put(3.8,-1){\huge{h}}
\put(4.2,-1.2){\bf{\Large{3}}}
\end{picture}
\vspace{6cm}
\caption{An up spin followed by a doublet. Site-1 may have flipped up 
under a $p_0$-process or a $p_1$-process. Subsequently when site-3 
flips up we get a singlet at site-2. The fraction of such singlets 
depends on the details of how sites 1 and 3 have flipped.}
\label{fig 14}
\end{figure}

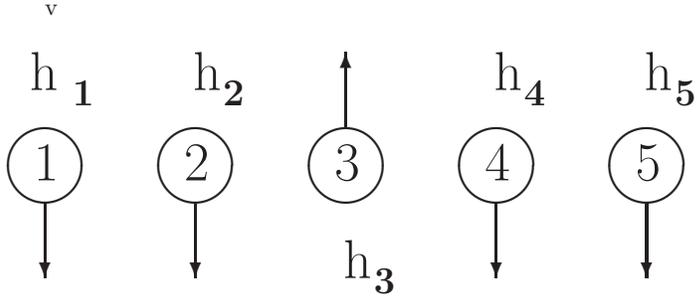
\begin{figure}
\setlength{\unitlength}{1cm}
\begin{picture}(9,0.)
\thicklines
\put(0,-2){\circle{1}}
\put(-.15,-2.2){\huge{1}} \put(0,-2.5){\vector(0,-1){1}}
\put(2,-2){\circle{1}}
\put(1.85,-2.2){\huge{2}} \put(2,-2.5){\vector(0,-1){1}}
\put(4,-2){\circle{1}}
\put(3.85,-2.2){\huge{3}} \put(4,-1.5){\vector(0,1){1}}
\put(6,-2){\circle{1}}
\put(5.85,-2.2){\huge{4}} \put(6,-2.5){\vector(0,-1){1}}
\put(8,-2){\circle{1}}
\put(7.85,-2.2){\huge{5}} \put(8,-2.5){\vector(0,-1){1}}
\put(-0.2,-1){\huge{h}}
v\put(0.2,-1.2){\bf{\Large{1}}}
\put(1.8,-1){\huge{h}}
\put(2.2,-1.2){\bf{\Large{2}}}
\put(3.8,-3.5){\huge{h}}
\put(4.2,-3.7){\bf{\Large{3}}}
\put(5.8,-1){\huge{h}}
\put(6.2,-1.2){\bf{\Large{4}}}
\put(7.8,-1){\huge{h}}
\put(8.2,-1.2){\bf{\Large{5}}}
\end{picture}
\vspace{6cm}
\caption{ A doublet followed by a doublet. We get a singlet at site-3 if 
sites 2 and 4 flip up before sites 1 and 5 and site-3 flips down because 
it is unstable with both neighbors up.}
\label{fig 15}
\end{figure}
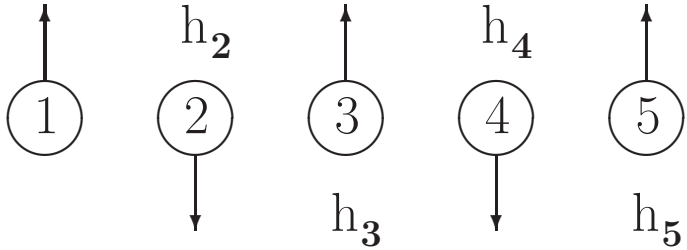
\begin{figure}
\setlength{\unitlength}{1cm}
\begin{picture}(9,0.)
\thicklines
\put(0,-2){\circle{1}}
\put(-.15,-2.2){\huge{1}} \put(0,-1.5){\vector(0,1){1}}
\put(2,-2){\circle{1}}
\put(1.85,-2.2){\huge{2}} \put(2,-2.5){\vector(0,-1){1}}
\put(4,-2){\circle{1}}
\put(3.85,-2.2){\huge{3}} \put(4,-1.5){\vector(0,1){1}}
\put(6,-2){\circle{1}}
\put(5.85,-2.2){\huge{4}} \put(6,-2.5){\vector(0,-1){1}}
\put(8,-2){\circle{1}}
\put(7.85,-2.2){\huge{5}} \put(8,-1.5){\vector(0,1){1}}
\put(1.8,-1){\huge{h}}
\put(2.2,-1.2){\bf{\Large{2}}}
\put(3.8,-3.5){\huge{h}}
\put(4.2,-3.7){\bf{\Large{3}}}
\put(5.8,-1){\huge{h}}
\put(6.2,-1.2){\bf{\Large{4}}}
\put(7.8,-3.5){\huge{h}}
\put(8.2,-3.7){\bf{\Large{5}}}
\end{picture}
\vspace{6cm}
\caption{ A singlet followed by a singlet. A singlet is created at 
site-3 if sites 2 and 4 flip up under a $p_2$-process and then site-3 
flips down because it is unstable when both neighbors are up.}
\label{fig 16}
\end{figure}
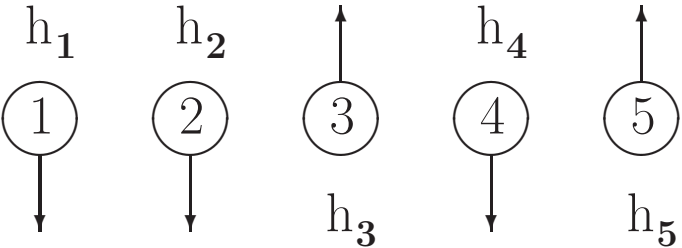
\begin{figure}
\setlength{\unitlength}{1cm}
\begin{picture}(9,0.)
\thicklines
\put(0,-2){\circle{1}}
\put(-.15,-2.2){\huge{1}} \put(0,-2.5){\vector(0,-1){1}}
\put(2,-2){\circle{1}}
\put(1.85,-2.2){\huge{2}} \put(2,-2.5){\vector(0,-1){1}}
\put(4,-2){\circle{1}}
\put(3.85,-2.2){\huge{3}} \put(4,-1.5){\vector(0,1){1}}
\put(6,-2){\circle{1}}
\put(5.85,-2.2){\huge{4}} \put(6,-2.5){\vector(0,-1){1}}
\put(8,-2){\circle{1}}
\put(7.85,-2.2){\huge{5}} \put(8,-1.5){\vector(0,1){1}}
\put(-0.2,-1){\huge{h}}
\put(0.2,-1.2){\bf{\Large{1}}}
\put(1.8,-1){\huge{h}}
\put(2.2,-1.2){\bf{\Large{2}}}
\put(3.8,-3.5){\huge{h}}
\put(4.2,-3.7){\bf{\Large{3}}}
\put(5.8,-1){\huge{h}}
\put(6.2,-1.2){\bf{\Large{4}}}
\put(7.8,-3.5){\huge{h}}
\put(8.2,-3.7){\bf{\Large{5}}}
\end{picture}
\vspace{6cm}
\caption{ A doublet followed by a singlet. We get a singlet at site-3 if 
site-2 flips up under a $p_1$-process, site-4 flips up under a 
$p_2$-process and then site-3 flips down because it is unstable when 
both neighbors are up.}
\label{fig 17}
\end{figure}

\begin{figure}[p] 
\includegraphics[width=.75\textwidth,angle=0]{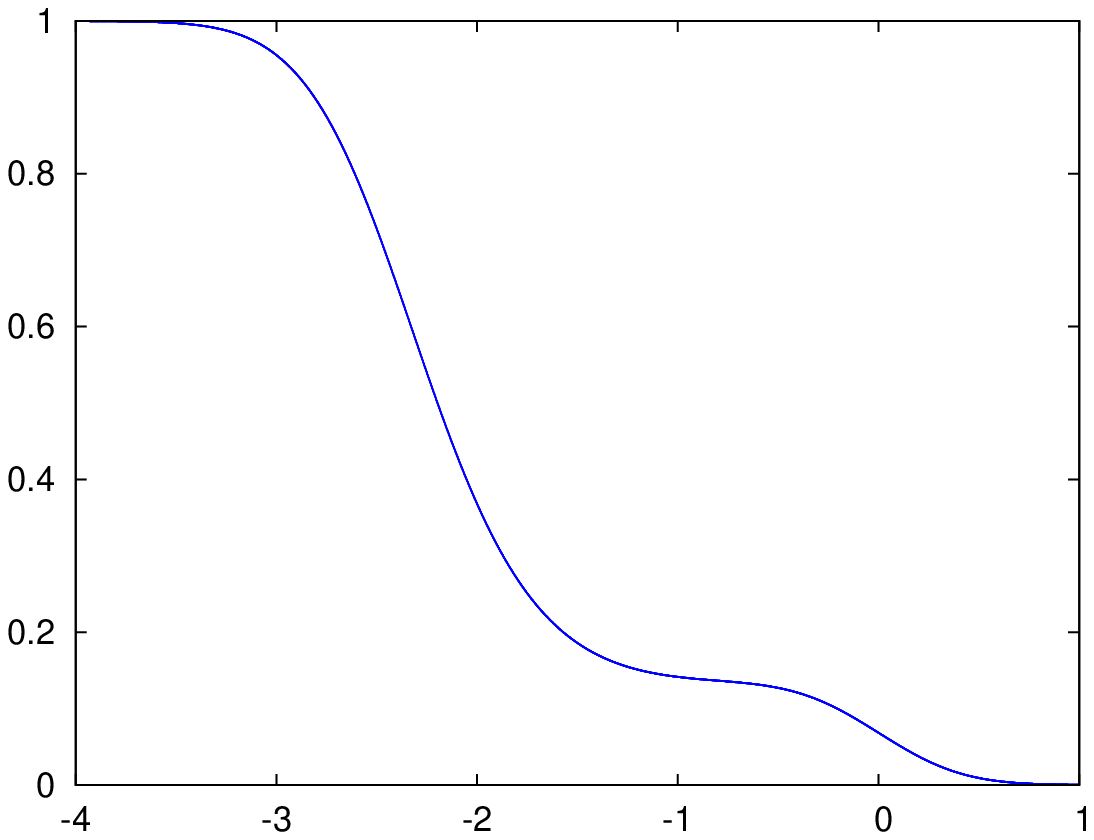} \caption{ 
Probability of a doublet $P_{\downarrow\downarrow}(h_a)$ for a Gaussian 
distribution with $\sigma=0.5 |J|$. Simulation data has been 
superimposed on the theoretical expression.} \label{fig 18} \end{figure}

\begin{figure}[p] 
\includegraphics[width=.75\textwidth,angle=0]{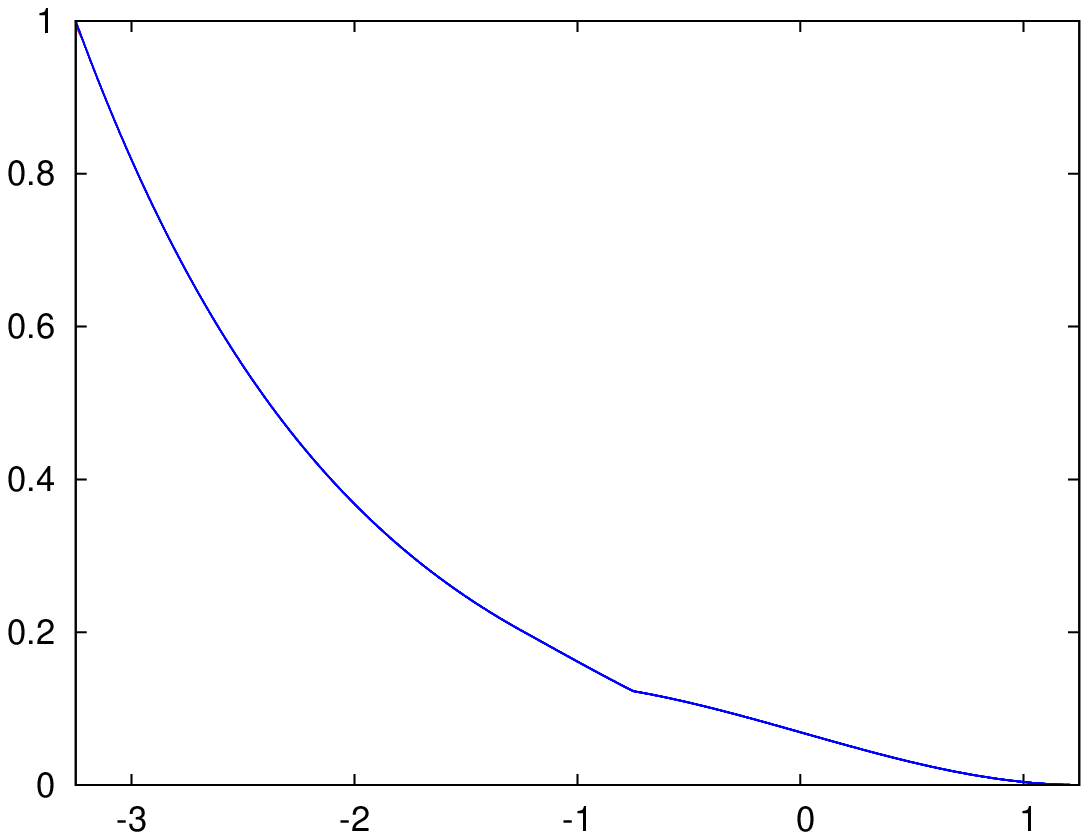} \caption{ 
Probability of a doublet $P_{\downarrow\downarrow}(h_a)$ for a uniform 
distribution with $\Delta=1.25 |J|$. Simulation data has been 
superimposed on the theoretical expression.} \label{fig 19} \end{figure}

\begin{figure}[p] 
\includegraphics[width=.75\textwidth,angle=0]{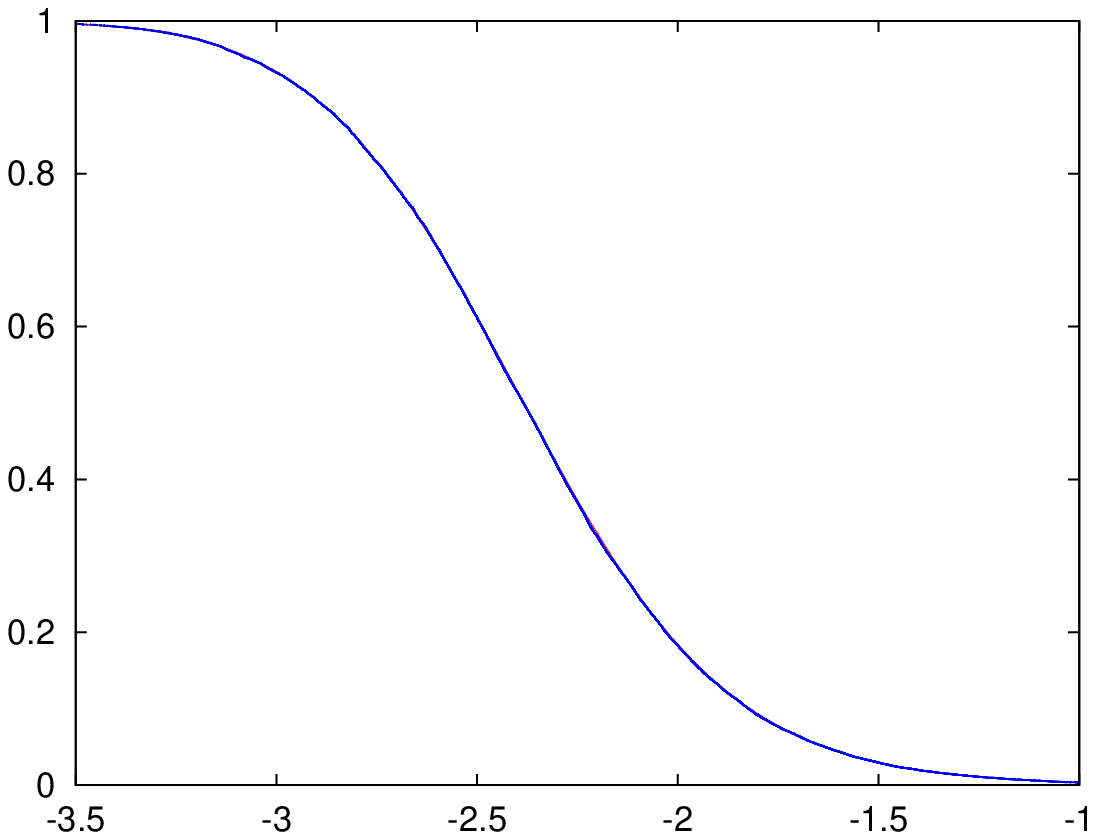} \caption{ 
Probability of $P_{\downarrow\downarrow\downarrow}(h_a)$ for a Gaussian 
distribution with $\sigma=0.5 |J|$. Simulation data has been 
superimposed on the theoretical expression.} \label{fig 20} \end{figure}

\begin{figure}[p] 
\includegraphics[width=.75\textwidth,angle=0]{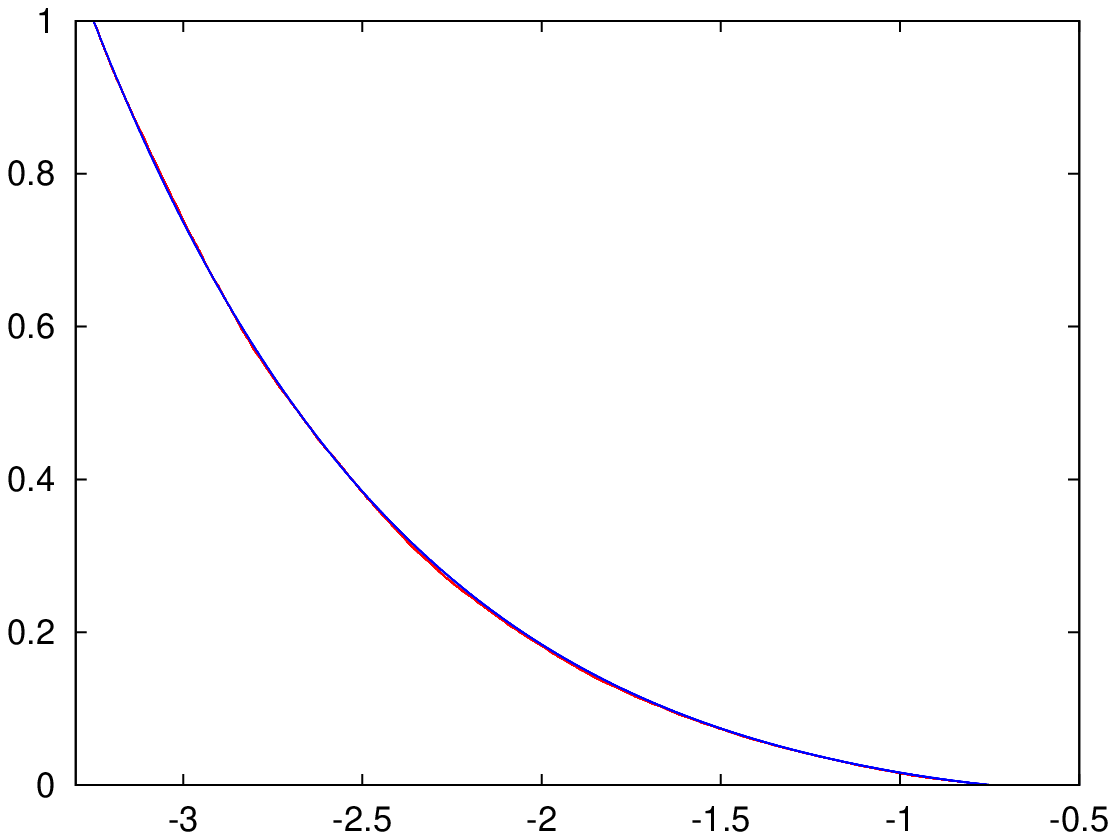} \caption{ 
Probability of $P_{\downarrow\downarrow\downarrow}(h_a)$ for a uniform 
distribution with $\Delta=1.25 |J|$. Simulation data has been 
superimposed on the theoretical expression.} \label{fig 21} \end{figure}

\begin{figure}[p] 
\includegraphics[width=.75\textwidth,angle=0]{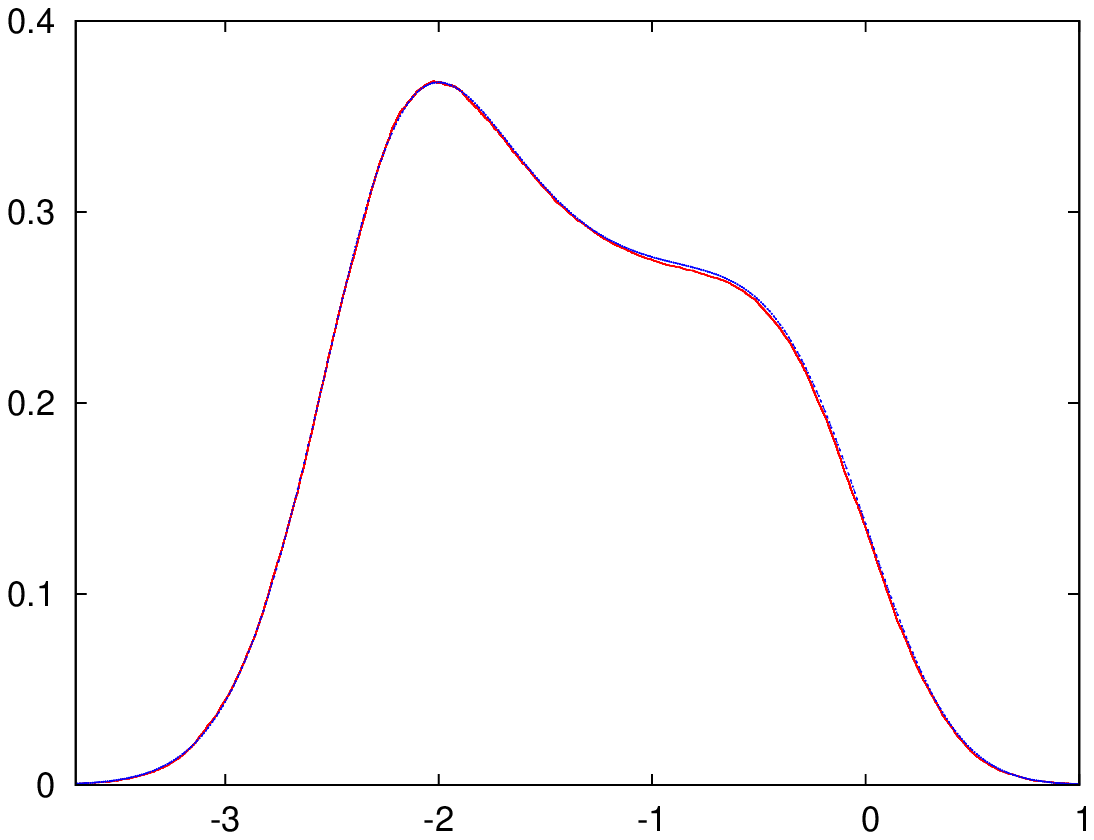} \caption{ 
$P_{\uparrow\downarrow\downarrow}(h_a) + 
P_{\downarrow\downarrow\uparrow}(h_a)$ for a Gaussian distribution with 
$\sigma=0.5 |J|$. Simulation data has been superimposed on the 
theoretical expression.} \label{fig 22} \end{figure}

\begin{figure}[p] 
\includegraphics[width=.75\textwidth,angle=0]{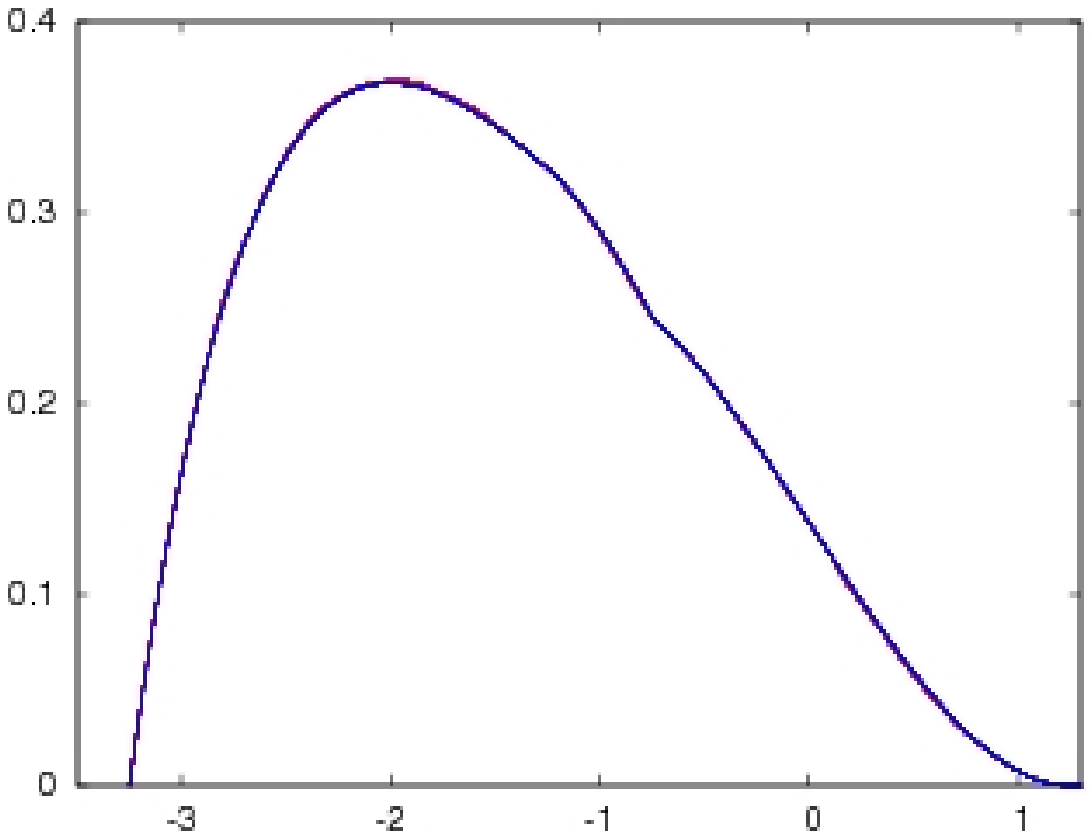} \caption{ 
$P_{\uparrow\downarrow\downarrow}(h_a) + 
P_{\downarrow\downarrow\uparrow}(h_a)$ for a uniform distribution with 
$\Delta=1.25 |J|$. Simulation data has been superimposed on the 
theoretical expression.} \label{fig 23} \end{figure}

\begin{figure}[p] 
\includegraphics[width=.75\textwidth,angle=0]{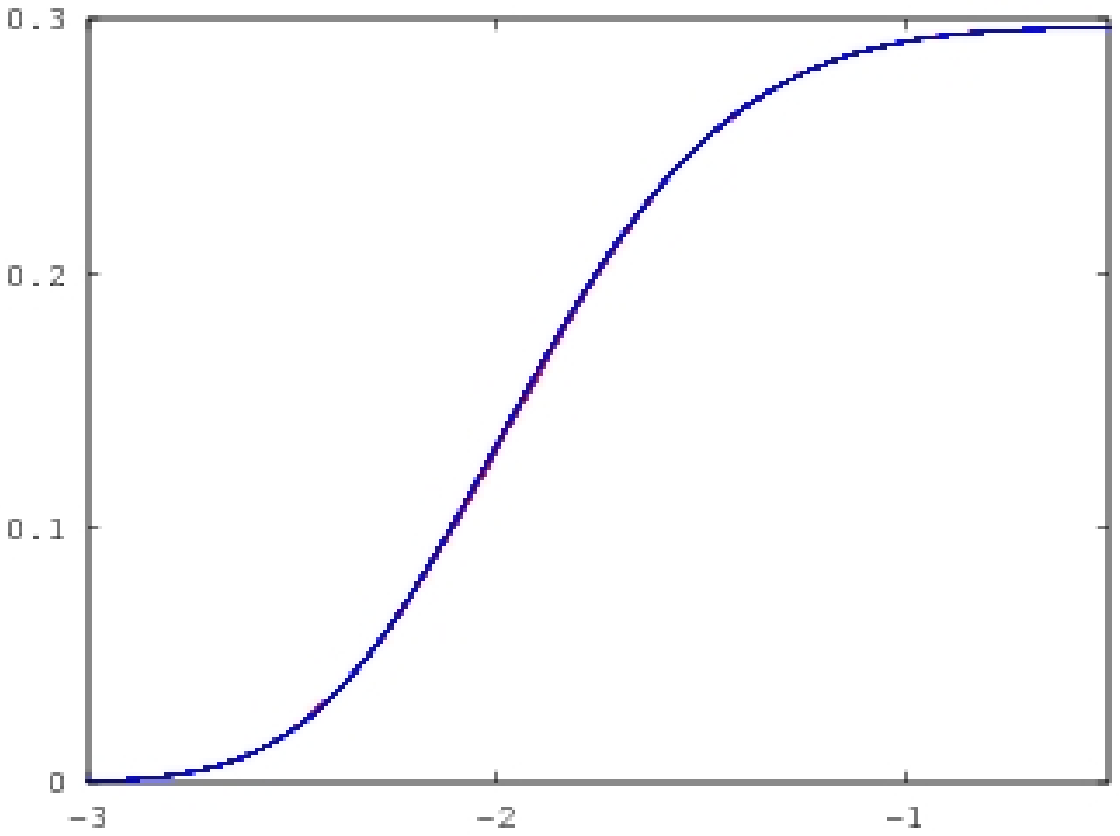} \caption{ 
Theory and simulation for $P_{\uparrow\downarrow\uparrow A}(h_a)$ for a 
Gaussian distribution with $\sigma=0.5 |J|$.} \label{fig 24} 
\end{figure}

\begin{figure}[p] 
\includegraphics[width=.75\textwidth,angle=0]{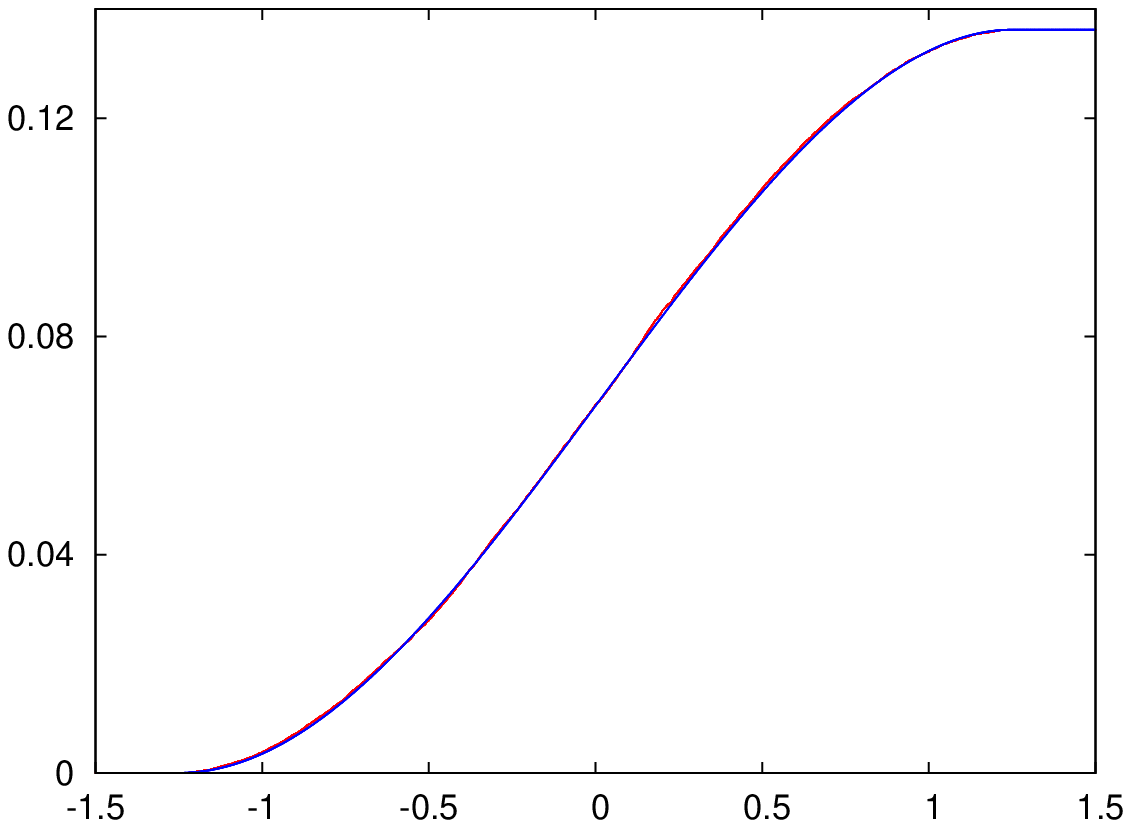} \caption{ 
Theory and simulation for $P_{\uparrow\downarrow\uparrow D}(h_a)$ for a 
uniform distribution with $\Delta=1.25 |J|$.} \label{fig 25} 
\end{figure}

\begin{figure}[p] 
\includegraphics[width=.75\textwidth,angle=0]{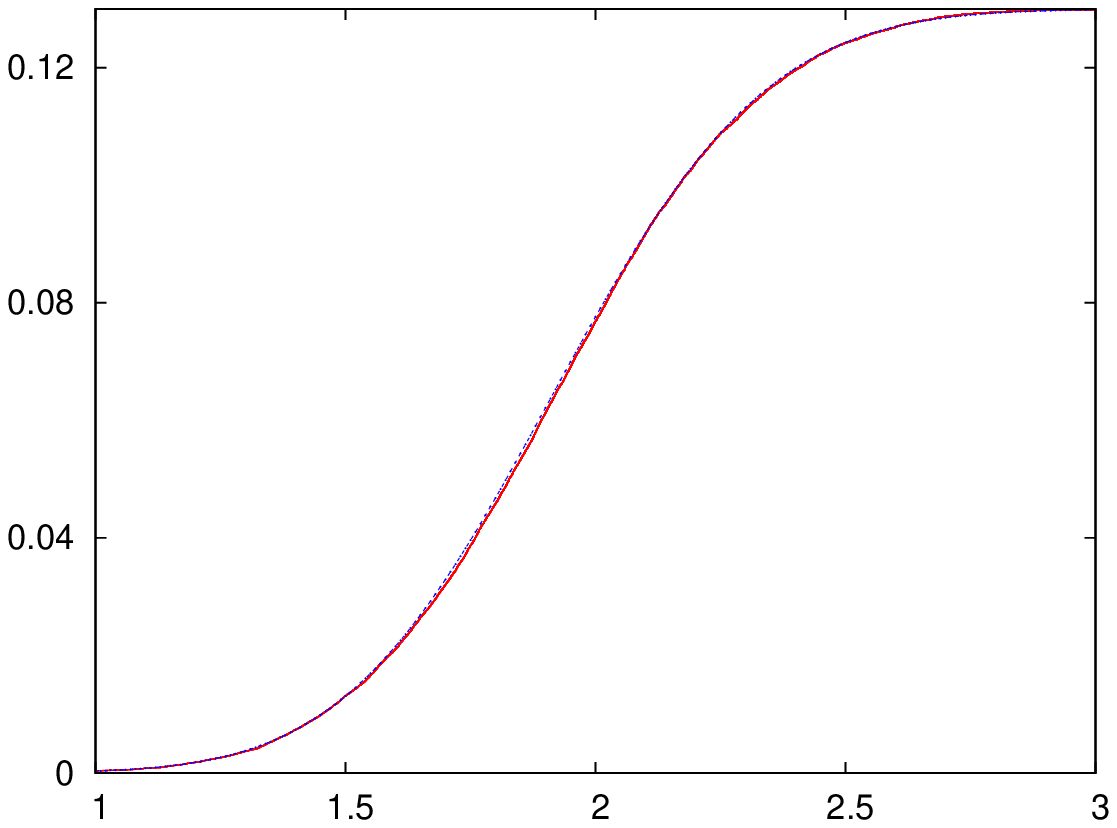} \caption{ 
Theory and simulation for $P_{\uparrow\uparrow\uparrow M}(h_a)$ for a 
Gaussian distribution with $\sigma=0.5 |J|$.} \label{fig 26} 
\end{figure}

\begin{figure}[p] 
\includegraphics[width=.75\textwidth,angle=0]{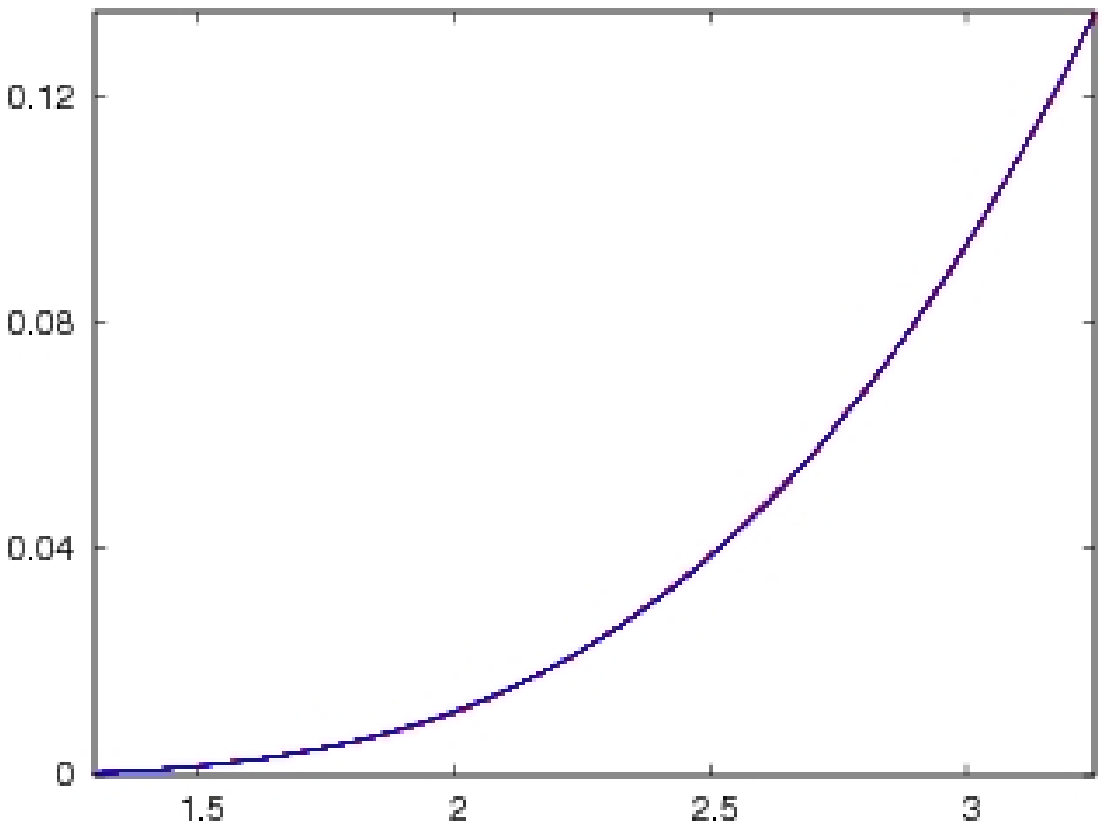} \caption{ 
Theory and simulation for $P_{\uparrow\uparrow\uparrow N}(h_a)$ for a 
uniform distribution with $\Delta=1.25 |J|$.} \label{fig 27} 
\end{figure}

\begin{figure}[p] 
\includegraphics[width=.75\textwidth,angle=0]{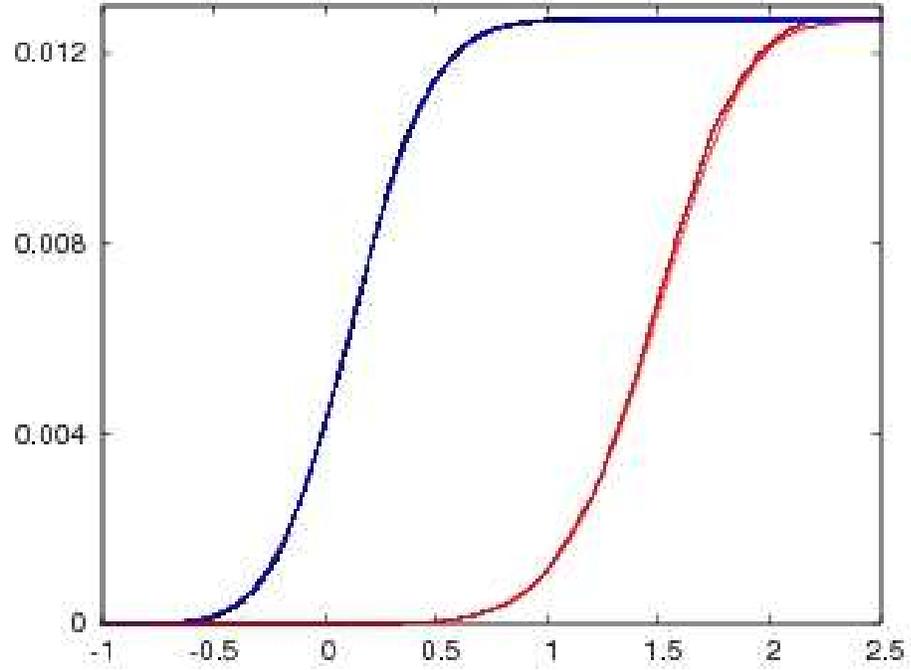} \caption{ 
Theory and simulation for $P_{\uparrow\downarrow\uparrow Q}(h_a)$ on the 
left, and $P_{\uparrow\uparrow\uparrow U}(h_a)$ on the right for a 
Gaussian distribution with $\sigma=0.5 |J|$.} \label{fig28} \end{figure}

\begin{figure}[p] 
\includegraphics[width=.75\textwidth,angle=0]{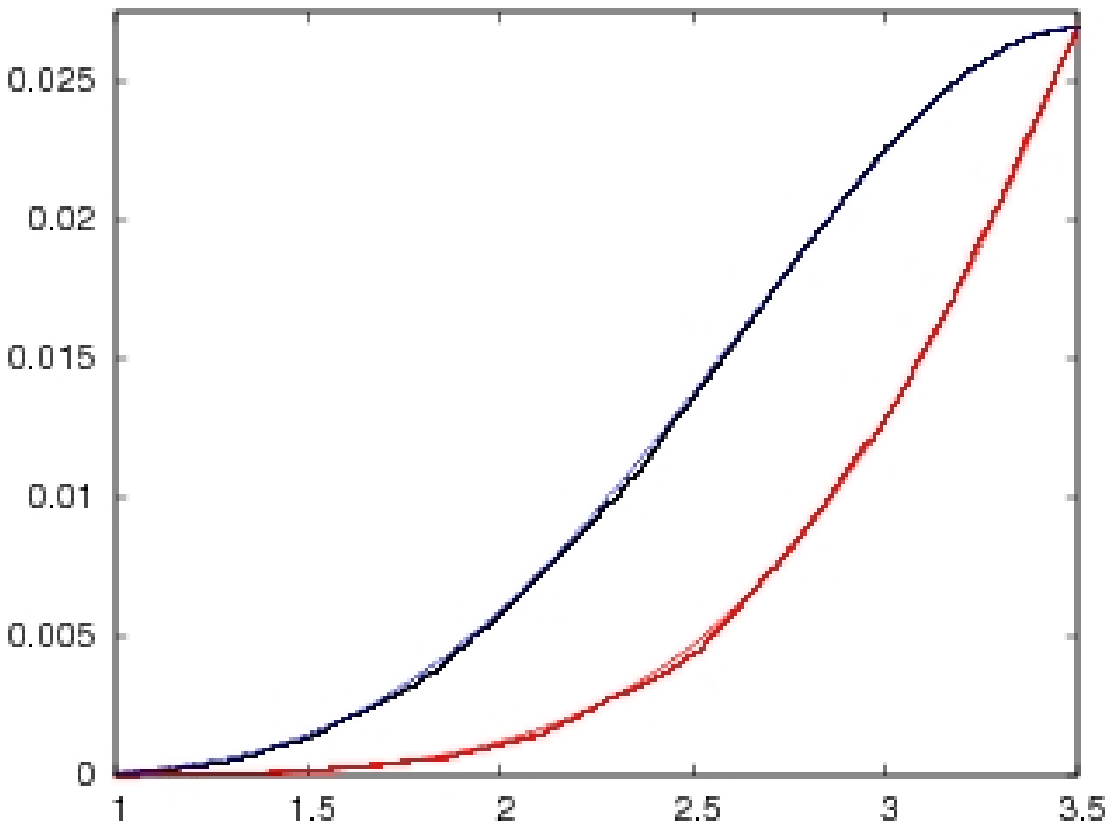} \caption{ 
Theory and simulation for $P_{\uparrow\downarrow\uparrow R}(h_a)$ on the 
left, and $P_{\uparrow\uparrow\uparrow V}(h_a)$ on the right for a 
uniform distribution with $\Delta=1.25 |J|$.} \label{fig 29} 
\end{figure}


\begin{thebibliography}{99}

\bibitem{bertotti} See for example, {\em{The Science of Hysteresis}}, 
edited by G Bertotti and I Mayergoyz, Academic Press, Amsterdam (2006).

\bibitem{young} {\em{Spinglasses and Random Fields}}, ed. A P 
Young, World Scientific, Singapore (1997).

\bibitem{sethna1}J P Sethna, K A Dahmen, S Kartha, J A 
Krumhansl, B W Roberts, and J D Shore, Phys Rev Lett 70, 3347 
(1993); K Dahmen and J P Sethna, Phys Rev Lett 71, 
3222(1993); O Perkovic, K Dahmen, and J P Sethna, Phys Rev 
Lett 75, 4528 (1995).

\bibitem{sethna2} J P Sethna, K A Dahmen, and O. Percovic in 
~\cite{bertotti} and references therein.

\bibitem{dahmen} K A Dahmen and J P Sethna, Phys Rev B 53, 14872 (1996).

\bibitem{imryma} Y Imry and S Ma Phys Rev Lett 35, 1399 (1975); for a 
more recent review of random field phenomena see e.g. T Nattermann in 
~\cite{young}.

\bibitem{glauber} R J Glauber, J Math Phys 4, 294 (1963).

\bibitem{wilson} K G Wilson, Phys Rev B4, 3174(1971); K G Wison, Phys 
Rev B4, 3184 (1971).

\bibitem{stoner} See for example, E C Stoner, Rev Mod Phys 25, 2 (1953). 
The main features of Barkhausen noise discovered in these early 
experiments are still topics of current research.

\bibitem{shukla1} Hysteresis in an Ising chain with quenched random 
disorder, P Shukla, Prog Theo Phys 96, 69-80 (1996).

\bibitem{shukla2} Exact solution of zero-temperature hysteresis in a 
ferromagnetic Ising chain with quenched random fields, P Shukla, Physica 
A233, 235-241 (1996).

\bibitem{dhar} D Dhar, P Shukla, and J P Sethna, J Phys A30, 5259 
(1997).

\bibitem{sabhapandit1} S Sabhapandit, D Dhar, and P Shukla, Phys Rev 
Lett 88, 197202 (2002).

\bibitem{silveira1} R da Silveira and M Kardar, Phys Rev E 59, 1355 
(1999).

\bibitem{silveira2} R A da Silveira and S Zapperi, Phys Rev B 69, 212404 
(2004).

\bibitem{shukla3} Hysteresis in random-field XY and Heisenberg models: 
Mean-field theory and simulations at zero temperature, Prabodh Shukla 
and R S Kharwanlang, Phys Rev E 81, 031106 (2010); cond-mat 
arXiv:0907.1957v2

\bibitem{shukla4} Critical hysteresis in random-field XY and Heisenberg 
models, Prabodh Shukla and R S Kharwanlang, Phys Rev E 83, 011121 
(2011); cond-mat arXiv:1007.4265v2

\bibitem{shukla5} Zero-temperature hysteresis in an anti-ferromagnetic 
Ising chain with quenched random fields, P Shukla, Physica A233, 242-252 
(1996).

\bibitem{shukla6} Hysteretic response of an anti-ferromagnetic random 
field Ising model in one dimension at zero temperature, Prabodh Shukla, 
Ratnadeep Roy, and Emilia Ray, Physica A 275, 380 (2000); 
cond-mat/9910021.

\bibitem{shukla7}item Hysteresis in one-dimensional anti-ferromagnetic 
random field Ising model at zero temperature, Prabodh Shukla, Ratnadeep 
Roy, and Emilia Ray, Physica A 276, 365 (2000); cond-mat/9910169.

\bibitem{shukla8} Driven Random Field Ising Model: some exactly solved 
examples in threshold activated kinetics, Prabodh Shukla, Int J Mod Phys 
B 17, 5583 (2003); cond-mat/0401252.

\bibitem{shukla9} Exact solution of return hysteresis loops in a one 
dimensional random-field Ising model at zero temperature, Prabodh 
shukla, Phys Rev E 62, 4725 (2000); cond-mat/0004125.

\bibitem{shukla10} Exact expressions for minor hysteresis loops in the 
random-field Ising model on a Bethe lattice at zero temperature, Prabodh 
Shukla, Phys Rev E 63, 27102 (2001);\\ cond-mat/0007370.

\bibitem{sabhapandit2} S Sabhapandit, P Shukla, and D Dhar, J Stat Phys 
98, 103 (2000).


\bibitem{illa1} X Illa, P Shukla, and E Vives, Phys Rev B 73, 092414 
(2006).

\bibitem{illa2} X Illa, M L Rosinberg, P Shukla, and E Vives, Phys Rev B 
74, 224404 (2006).

\bibitem{illa3} X Illa, M L Rosinberg, and G Tarjus, Eur Phys J B54, 355 
(2006).

\bibitem{ayyub} P Ayyub et al, Phys Rev B 57, R5559 (1998).

\bibitem{fukuma} K Fukuma and M Torii, Earth Planets Space 50, 9 (1998).

\bibitem{chiorescu} I Chiorescu et al, Phys Rev Lett 84, 3454 (2000).

\bibitem{fullerton} E E Fullerton et al, Appld Phys Lett 77, 3806(2000).

\bibitem{waldmann} O Waldmann et al, Phys Rev Lett 89, 246401(2002).

\bibitem{takanashi} K Takanashi, Appld Phys Lett 63, 1585 (1993).

\bibitem{chew} Khian-Hooi Chew et al, Appld Phys Lett 77, 2755 (2000).

\bibitem{kisker} J Kisker,H Rieger, and H Schreckenberg, J Phys A: Math. 
Gen. 27, L853 (1994). This paper discusses the non-equilibrium dynamics 
of a non-random, one-dimensional Ising model, with three-spin 
interactions, at low temperatures. It is qualitatively similar to the 
dynamics of the random field Ising model at zero-temperature.

\bibitem{rieger} H Reiger, Physica A 224, 267 (1996).

\bibitem{ritort} F Ritort and P Sollich, Advances in Physics 52:4, 
219 (2003).

\bibitem{toninelli} C Toninelli, PhD thesis (2003).

\bibitem{evans} J W Evans, Rev Mod Phys 65, 1281 (1993).

\bibitem{mityushin} L G Mityushin, Prob Peredachi Inf 9, 81 (1973). Also 
see reference (39) for a detailed discussion of the screening property 
of this class of problems.

\end{thebibliography}
\end{document}